\documentclass[5p]{elsarticle}

\usepackage{booktabs} 
\usepackage{listings}
\usepackage{changepage}
\usepackage{url}
\usepackage{interval}
\usepackage{caption}
\usepackage{subcaption}

\usepackage{amssymb} 

\usepackage{amsmath}
\usepackage[artemisia]{textgreek} 

\usepackage{xcolor}

\usepackage[ruled,linesnumbered]{algorithm2e} 

\SetAlFnt{\small}
\SetAlCapFnt{\small}
\SetAlCapNameFnt{\small}
\SetAlCapHSkip{0pt}
\IncMargin{-\parindent}

\def\eg{e.\,g.}
\def\ie{i.\,e.}
\newcommand{\mli}[1]{\mathit{#1}}

\begin{document}
\begin{frontmatter}

\title{Using hardware performance counters to speed up autotuning convergence on GPUs}

\author[ics]{Ji\v{r}\'{i} Filipovi\v{c}\corref{mycorrespondingauthor}}
\ead{fila@mail.muni.cz}
\cortext[mycorrespondingauthor]{Corresponding author}
\author[ics]{Jana Hozzov\'{a}}
\ead{hozzova@mail.muni.cz}
\author[ics]{Amin Nezarat}
\ead{aminnezarat@mail.muni.cz}
\author[ics]{Jaroslav Ol'ha}
\ead{348646@mail.muni.cz}
\author[ics]{Filip Petrovi\v{c}}
\ead{fillo@mail.muni.cz}
\address[ics]{Institute of Computer Science, Masaryk University, Botanick\'{a} 68a, 60200 Brno, Czech Republic}

\begin{abstract}
Nowadays, GPU accelerators are commonly used to speed up general-purpose computing tasks on a variety of hardware. However, due to the diversity of GPU architectures and processed data, optimization of codes for a particular type of hardware and specific data characteristics can be extremely challenging. The autotuning of performance-relevant source-code parameters allows for automatic optimization of applications and keeps their performance portable. Although the autotuning process typically results in code speed-up, searching the tuning space can bring unacceptable overhead if (i) the tuning space is vast and full of poorly-performing implementations, or (ii) the autotuning process has to be repeated frequently because of changes in processed data or migration to different hardware.

In this paper, we introduce a novel method for searching generic tuning spaces. The tuning spaces can contain tuning parameters changing any user-defined property of the source code. The method takes advantage of collecting hardware performance counters (also known as profiling counters) during empirical tuning. Those counters are used to navigate the searching process towards faster implementations. The method requires the tuning space to be sampled on any GPU. It builds a problem-specific model, which can be used during autotuning on various, even previously unseen inputs or GPUs. Using a set of five benchmarks, we experimentally demonstrate that our method can speed up autotuning when an application needs to be ported to different hardware or when it needs to process data with different characteristics. We also compared our method to state of the art and show that our method is superior in terms of the number of searching steps and typically outperforms other searches in terms of convergence time.
\end{abstract}



\end{frontmatter}	

\section{Introduction}

In the recent decade, GPU accelerators have been used to speed up non-graphical applications on multiple classes of devices -- from portable devices to supercomputers. Different models of GPU hardware vary significantly in terms of their architecture, even when manufactured by the same vendor. For example, there are eight generations of NVIDIA CUDA GPUs available, differing in the number of cores per multiprocessors, the presence of read/write L1 cache, the number of registers and more. Moreover, each generation contains a broad range of GPU models with different floating-point performance and memory bandwidth. As the hardware characteristics heavily influence the performance of a given GPU kernel, its code needs to be adapted for each GPU model to achieve optimal performance -- otherwise, performance portability is not ensured~\cite{kurzak2012autotuning, nugteren2015cltune, petrovic2020benchmark}. Furthermore, kernels' performance is also sensitive to input size, structure, or application settings, so a code optimized for a certain input characteristics may run sub-optimally when those change~\cite{gonzalo2017revisiting, nugteren2018clblast, strelak2019gpu}.

Autotuning is a method which allows codes to be automatically adjusted to given hardware and input~\cite{balaprakash2018autotuning}. During GPU kernel development, programmers define \textit{tuning parameters} -- the properties of the code which influence application performance. Each tuning parameter can take one of a pre-defined set of discrete values. The cross product of tuning parameters (potentially pruned by \textit{a priori} known constraints) forms a \textit{tuning space}. One point in tuning space, which defines how the computational kernel is created and executed, is called a \textit{tuning configuration}. An autotuning framework searches the tuning space for a tuning configuration that minimizes a tuning objective, usually runtime or power consumption~\cite{balaprakash2018autotuning}. 

Autotuning spaces have several properties that make their efficient search difficult. These discrete optimization spaces with many dimensions are known to be non-convex, non-linear and with low locality~\cite{balaprakash2011can}. The time needed to perform a tuning space search can limit the practical usage of autotuning. This happens especially in two cases: (i) when the tuning space is vast, and most of its configurations perform poorly, thus the search takes long; (ii) when the performance depends significantly on input characteristics (\eg{}, size), which often change, thus the search needs to happen often.

Methods for searching tuning spaces view their objective as a function of tuning parameters. Multiple methods for tuning specific optimization exist~\cite{cummins2017end, connors2019automatically, yu2020efficient, muralidharan2016architecture}, some of them leveraging performance counters~\cite{wang2009mapping, singh2009prediction, cavazos2007rapidly, rahman2015maximizing, zhang2018auto}. Those methods are designed for fixed types of optimizations (\ie{}, tuning parameters) and can tune those parameters for unseen applications on the seen hardware. It is possible because they use a model for concrete optimization, so the model can be constructed using a base of many applications implementing the same type of optimization. 

In contrast, generic tuning spaces can contain any optimization the programmer implements, so the model cannot be constructed this way. State-of-the-art methods for searching generic tuning spaces (\ie{}, including any tuning parameters) are based on mathematical optimization~\cite{balaprakash2011can, vanwerkhoven2018kernel}, or they use a surrogate performance/power model built from a sample of the tuning space~\cite{jia2013starchart, price2015improving, falch2015machine, feng2017sampling}. Because the function relating tuning parameters with the tuning objective differs with hardware and input, those methods require the autotuning to be repeated from scratch when hardware or input changes.

The essential contribution of this paper is the introduction of a novel generic tuning space search method that breaks the aforementioned relation. Once the tuning space is partially explored for some hardware and input, the portable model makes it possible to speed up tuning when input or hardware changes. 
Our method mimics the iterative optimization performed by developers. Developers use profilers to collect performance counters and identify bottlenecks (overloaded processor subsystems) on multiple levels of hardware hierarchy. They also understand how the properties of their code (\ie{}, tuning parameters) are related to performance counters. They iteratively detect bottlenecks and modify the code to reduce stress on the bottlenecks until they reach sufficient code performance. Our method aims to do the same. It requires a training phase, wherein the model is trained to capture how tuning parameters influence performance counters (machine-learning based analogy to a developer's understanding of the relationship). During autotuning, the method iteratively profiles the code and acquires performance counters, analyzes bottlenecks and determines which performance counters should be changed to soften the bottlenecks using an expert system (analogy to a developer's work with a profiling tool). Then, it uses the model to determine which tuning configurations change performance counters in the required way. Finally, it selects the next tuning configuration to profile (analogy to a developer's modification of the code).

\begin{figure}[t]
    \centering
    \includegraphics[width=.49\hsize]{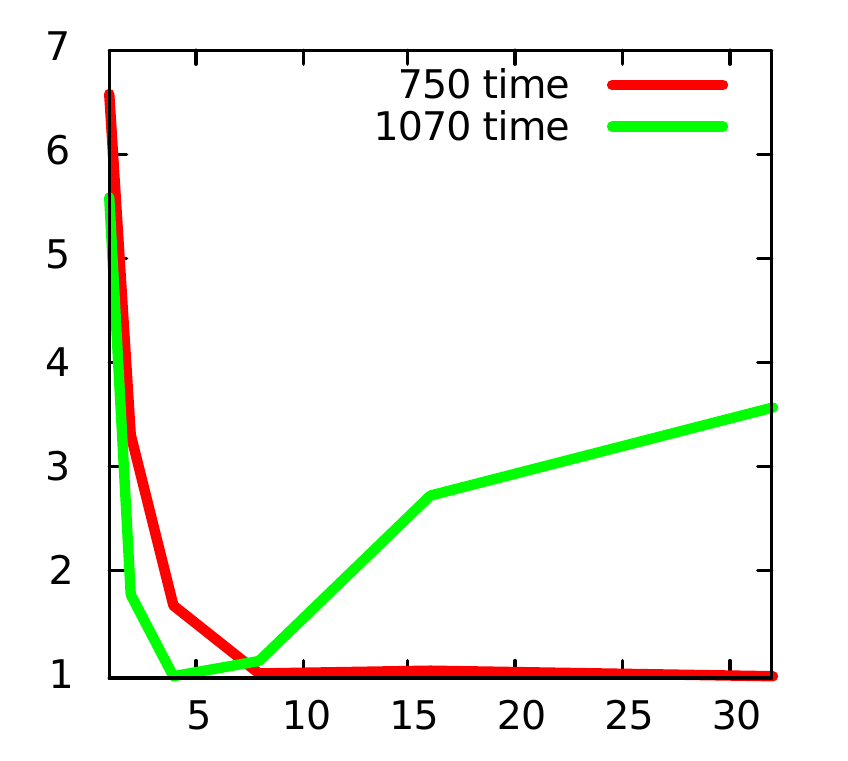}
    \includegraphics[width=.49\hsize]{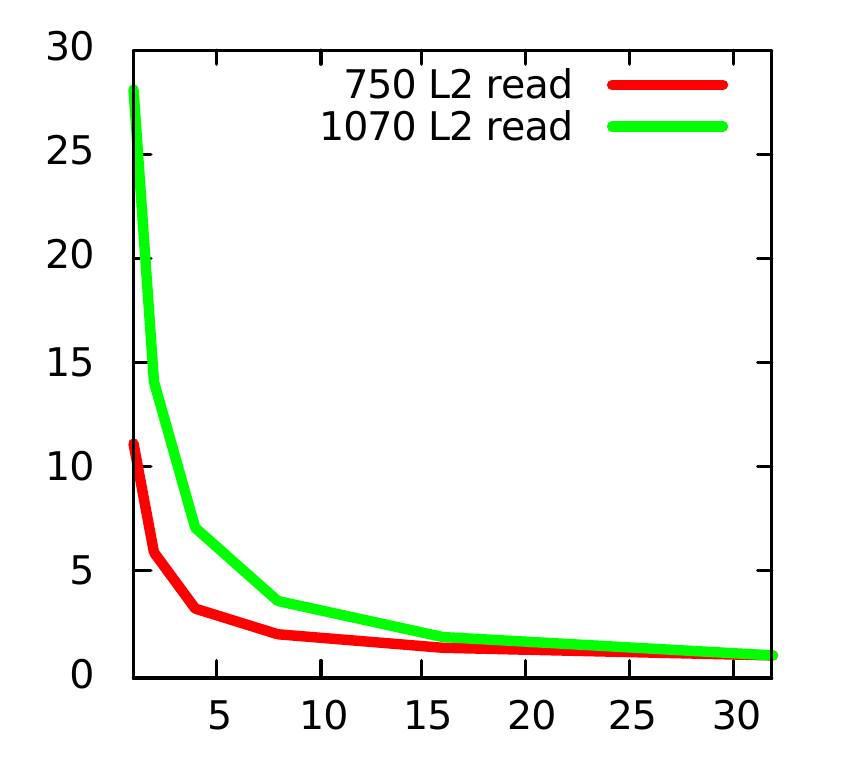}
    \includegraphics[width=.49\hsize]{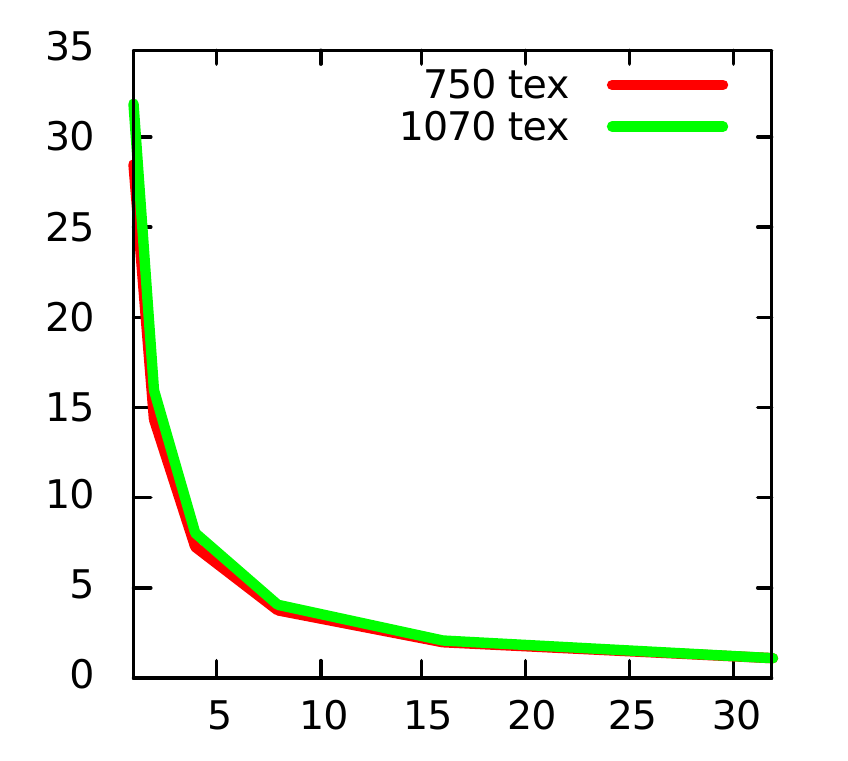}
    \includegraphics[width=.49\hsize]{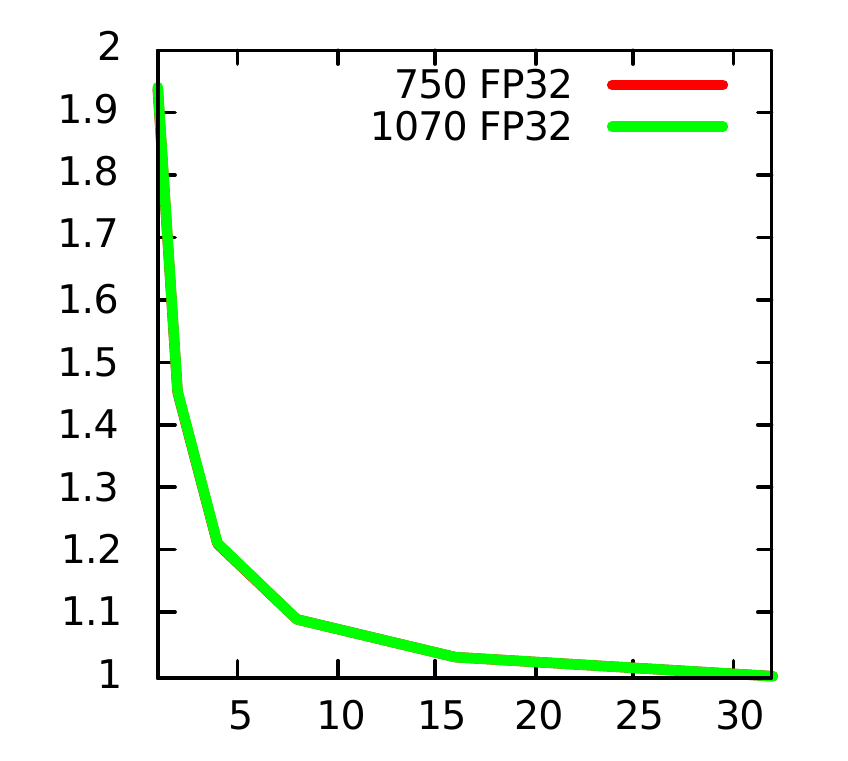}
    \caption{Dependence between a tuning parameter and various properties of the kernel, shown on Coulomb summation~\cite{filipovic2017autotuning} running with large gridbox on GeForce GTX 750 and with small gridbox on GeForce GTX 1070. The x-axis shows a tuning parameter changing thread coarsening. The y-axis shows normalized values of selected properties: kernel runtime, L2 cache read transactions, texture cache read transactions and 32-bit floating-point operations.}
    \label{fig:dependence-ilustration}
\end{figure}

The strength of our method is its ability to build a model using a particular GPU and input, and use this model to speed up autotuning of a kernel running on a \textit{different GPU} or processing \textit{different input}. 
This is possible because the method builds the model of relations between tuning parameters and performance counters and searches for the performance gain based on the performance counters, instead of relating tuning parameters directly to the performance, as is done in~\cite{balaprakash2011can, vanwerkhoven2018kernel, jia2013starchart, price2015improving, falch2015machine, feng2017sampling}. We distinguish two types of performance counters: the counters measuring stress of a processor subsystem, called $\mli{PC_{stress}}$, and counters measuring amount of operations on a processor subsystem, called $\mli{PC_{ops}}$ (assignment of selected NVIDIA GPU performance counters between those sets is shown in Table~\ref{tab:pcs}). Compared to the performance itself, the tuning parameters affect performance counters $\mli{PC_{ops}}$ in a more straightforward and stable way. 
This is illustrated in Figure~\ref{fig:dependence-ilustration} for the Coulomb sum benchmark (described in the next section of this paper). In this example, the relation of the depicted tuning parameter to kernel runtime changes significantly with different input and hardware, whereas the relation between the tuning parameter and normalized values of performance counters $\mli{PC_{ops}}$ remains stable~\footnote{On the other hand, relations between a given tuning parameter and performance counters $\mli{PC_{stress}}$, such as the utilization of caches or floating-point units, differ significantly between GPUs and input sizes.}. As we show in this paper, the stability of the relation between tuning parameters and performance counters when changing hardware or input makes it possible to train and use models created from historical tuning data to speed up tuning space search with various benchmarks. 

The evaluation is performed on a benchmark set of five kernels taken from~\cite{nugteren2015cltune, petrovic2020benchmark}. The obtained result shows that measuring the performance counters and using a model created from historical tuning data can bias the search process towards faster convergence. The proposed profile-based searcher systematically outperforms random search using a model created on different GPU, different input, or smaller tuning space. The proposed searcher also outperforms Basin Hopping implemented in Kernel Tuner~\cite{vanwerkhoven2018kernel} and regression-tree model implemented in Starchart~\cite{jia2013starchart} in most cases.

The paper makes the following major contributions:
\begin{itemize}
  \item \textit{Using hardware performance counters to navigate search\-ing of tuning space.} The proposed profile-based searcher is agnostic to the type of tuning parameters included in the tuning space (both their number and the code properties they tune). To the best of our knowledge, this is the first generic autotuning searcher using performance counters to speed up tuning space searching convergence with arbitrary tuning parameters that are unknown at the time of the searcher design.
  \item \textit{Comprehensive evaluation of the proposed profile-ba\-sed searcher.} The proposed searcher is tested in multiple scenarios, including performance portability across various hardware, input, and tuning spaces, using five benchmarks and four GPU architectures. We have also compared our searcher to an op\-ti\-mi\-za\-ti\-on-ba\-sed searcher~\cite{vanwerkhoven2018kernel} and a model-based searcher~\cite{jia2013starchart}, showing that the proposed searcher converges faster to the near-optimal configuration in most cases.
  \item \textit{Integration of the searcher into a real-world tuning framework.} The proposed profile-based searcher is implemented within Kernel Tuning Toolkit (KTT)~\cite{petrovic2020benchmark}. Therefore, it is possible to measure the real speed of searching convergence. Moreover, the tuning framework with the searcher and experimental data~\cite{filipovic2021cuda} is freely available to the community~\footnote{\url{https://github.com/HiPerCoRe/KTT}}.
\end{itemize}

The rest of the paper is organized as follows. In Section~\ref{sect:example}, an example of manual tuning using performance counters is given. This example serves to illustrate the process which our framework intends to automate. The proposed search method is described in Section~\ref{sect:method}. Section~\ref{sect:eval} evaluates the proposed searcher and compares it to alternative approaches. The related work is described in Section~\ref{sect:related_work}. We conclude and outline the future work in Section~\ref{sect:conclusion}.


\section{Example of Manual Tuning Space Search}
\label{sect:example}

In this section, we demonstrate a manual approach to tuning space search using hardware performance counters. This example aims to ease the understanding of the concepts behind our proposed automatic searcher described in the next section. 

\subsection{Direct Coulomb Summation}
We use a simplified version of the 3D implementation of Direct Coulomb Summation introduced in~\cite{filipovic2017autotuning} as an example. With Direct Coulomb Summation, the electrostatic potential around a molecule is computed on a regular grid. For each grid point $V_i$, we compute:
\begin{equation}
V_i = \sum_{j=0}^{n}\frac{w_j}{4 \pi \epsilon_0 r_{ij}}
\label{eq:coulomb}
\end{equation}
where $n$ is the number of atoms, $w_j$ is the charge of $j$-th atom, $r_{ij}$ is the Euclidean distance between atom $j$ and grid point $i$, and $\epsilon_0$ is vacuum permitivity. 

Listing~\ref{lst:coulomb} shows the source code of the simplified Direct Coulomb Summation kernel. The input of the kernel consists of \texttt{atomInfo}, containing atom coordinates and charges in a vector (\texttt{x}, \texttt{y}, \texttt{z} represent coordinates of the atom, \texttt{w} represents charge divided by $4 \pi \epsilon_0$), \texttt{numberOfAtoms} represents the number of atoms in \texttt{atomInfo}, \texttt{gridSpacing} is the size of a grid cell, \texttt{gridSize} is the number of grid cells in each dimension and \texttt{energyGrid} contains output potential energy grid. 

In our implementation, one GPU thread computes one or more grid points, determined by the value of \texttt{Z\_ITE\-RA\-TI\-ONS}, which is the only tuning parameter in this example. For each grid point, Equation~\ref{eq:coulomb} is computed. The code is expected to be compute-bound for two reasons: (i) for each atom's data loaded into the registers, we can compute values of multiple grid points, and (ii) all the threads within a warp load the same atom at the same time, so the code can benefit from good cache locality.

\begin{scriptsize}
\begin{lstlisting}[caption={Direct Coulomb Summation kernel},mathescape,escapeinside={(*}{*)}, label=lst:coulomb]
__global__ void directCoulombSum(
  const float4* atomInfo, int numberOfAtoms, 
  float gridSpacing, int gridSize, float* energyGrid)
{
  // integer coordinates within grid
  int x = blockIdx.x*blockDim.x + threadIdx.x;
  int y = blockIdx.y*blockDim.y + threadIdx.y;
  int z = (blockIdx.z*blockDim.z + threadIdx.z) 
    * Z_ITERATIONS;
        
  int slice = gridSize * gridSize;
  int out = slice*z + gridSize*y + x;

  // real coordinate within molecule space
  float fX = gridSpacing * xIndex;
  float fY = gridSpacing * yIndex;
  float fZ = gridSpacing * zIndex;

  // return value zeroizing
  float energyValue[Z_ITERATIONS];
  for (int i = 0; i < Z_ITERATIONS; i++) 
    energyValue[i] = 0.0f;

  // loop over all atoms 
  for (int i = 0; i < numberOfAtoms; i++)
  {
    float dX = fX - atomInfo[i].x;
    float dY = fY - atomInfo[i].y;
    float dZ = fZ - atomInfo[i].z;
    float w = atomInfo[i].w;
    // loop over multiple grid points
    for (int j = 0; j < Z_ITERATIONS; j++) {
      float rd = rsqrt(dX * dX + dY * dY + dZ*dZ);
      energyValue[j] += w * rd;
      dZ += gridSpacing;
    }
  }

  // store result in global memory
  for (int i = 0; i < Z_ITERATIONS; i++)
    if (z + i < gridSize)
      energyGrid[out + slice*i] += energyValue[i];
}
\end{lstlisting}
\end{scriptsize}

The code is autotuned by the KTT framework~\cite{petrovic2020benchmark}. KTT passes the tuning parameter via preprocessor macro \texttt{Z\_ITERATIONS}. For the sake of simplicity, we do not optimize thread block size in this example. When tuning the code, we will assign values from the set $\{1, 2, 4, 8, 16, 32\}$ to the tuning parameter.

\subsection{Understanding the effect of the tuning parameter}
Higher values of \texttt{Z\_ITERATIONS} improve data locality in registers: once atom coordinates and charge are read (lines 27-30), they are used multiple times in the loop beginning at line 32. Moreover, this parameter reduces invariant computation of $dX * dX + dY * dY$. However, a value of \texttt{Z\_ITERATIONS} that is too high reduces strong scaling (the kernel is executed by a lower amount of threads) and increases register consumption, which can lead to lower GPU occupancy or usage of slow local memory for register spilling. Knowing and understanding how tuning parameters relate to performance counters is part of a developer's expertise.

When profiling the code, the developer can set the value of the tuning parameter by observing hardware performance counters. In this example, several counters are highly relevant (some of them shown in Figure~\ref{fig:dependence-ilustration}):
\begin{itemize}
  \item utilization of floating point (FP) units: if high, \texttt{Z\_ITER\-ATIONS} should be set to a higher value (to reduce the number of FP operations by computing \texttt{dX * dX + dY * dY} fewer times);
  \item utilization of texture or L2 cache: if high (and not caused by register spilling), \texttt{Z\_ITERATIONS} should be set to a higher value (to improve cache locality);
  \item GPU occupancy: if low, \texttt{Z\_ITERATIONS} should be set to a lower value (to improve strong scaling);
  \item the amount of local memory: if high, \texttt{Z\_ITERATIONS} should be set to a lower value in case utilization of any part of the memory subsystem is also high (to decrease bandwidth introduced by register spills).
\end{itemize}

Knowing and understanding how performance counters relate to bottlenecks is also part of a developer's expertise. 

In this simple example, we only search a one-di\-men\-sion\-al tuning space. However, the rules mentioned here also apply with multiple tuning parameters: in such a case, the resulting kernel runtime and performance counter values are determined by a mixed effect of all tuning parameters, but \texttt{Z\_ITERATIONS} still affects performance counters in the same way.

\subsection{Tuning space search}
\label{sect:example_search}
Let us test our kernel using GeForce GTX 1070 with a grid of $256 \times 256 \times 256$ points and 64 atoms. The kernel should be compute-bound; therefore, we expect to have high utilization of FP units when tuned. Lets' set \texttt{Z\_ITERATIONS} to 1. When profiled, the kernel runtime is 10,475\,\textmu s. Performance counters point to a clear bottleneck in the texture cache (utilization level 9 out of 10), whereas FP instruction units are not highly loaded (utilization level 3 out of 10). Therefore, we raise the value of \texttt{Z\_ITERATIONS} to 32 to improve register locality. Now, the performance improves significantly: the kernel is executed in 1,531\,\textmu s. The performance counters show the following: texture cache utilization dropped from 9 to 3, FP utilization is 8, reported GPU occupancy is 0.6 (out of 1.0). We can try to increase occupancy to reduce pipeline and memory latencies. We set \texttt{Z\_ITERATIONS} to 8 and get a slightly better runtime of 1,497\,\textmu s. The GPU occupancy is now 0.98, FP utilization is 0.9 and texture cache utilization is 9. The texture cache utilization can be lowered by setting \texttt{Z\_ITERATIONS} to 16; however, this change does not lead to performance improvement; therefore, we can consider the kernel to be tuned.

Now, consider a change of the input data: using the same GPU, we will compute a grid of $25 \times 25 \times 25$ points and 4096 atoms. The previously tuned implementation, setting \texttt{Z\_ITERATIONS} to 8, yields a runtime of 394\,\textmu s. Occupancy of this implementation is 0.11. Some speedup can be reached by increasing occupancy, so we need to explore values lower than 8 for \texttt{Z\_ITERATIONS}. When set to 4, the runtime is improved to 260\,\textmu s. Occupancy is improved to 0.17, and we can already see the bottleneck on texture cache (utilization level 8), which suggests we should not decrease \texttt{Z\_ITERATIONS} further. When we try to set \texttt{Z\_ITERATIONS} to 2, the runtime increases to 428\,\textmu s. Although occupancy is improved to 0.32, the implementation is limited by the texture cache, which now reports utilization level 9. Therefore, we consider tuning to be finished, considering 4 to be the best value for \texttt{Z\_ITERATIONS}.

\subsection{Take-away from the example}
In this section, we showed how tuning is performed rationally by an expert programmer, who understands the relation between tuning parameters and performance counters, as well as the relation between performance counters and bottlenecks. The programmer profiles the tuned code and analyzes the observed bottlenecks. Then, they determine the direction of change of the tuning parameter based on an expected effect on performance counters. In the next section, we describe how to automatize the entire process.

\section{Proposed search method}
\label{sect:method}
In this section, we describe the basic idea and assumptions of the proposed profile-based search method, and then we focus on the details of our current implementation.

\subsection{On profiling counters}
\label{sect:pcs}
The tuning parameters (TPs) are tuned in order to optimize the code by balancing the usage of different processor subsystems. They change the properties of the code, thus lowering the stress on one processor subsystem and increasing the stress on another one\footnote{If any code transformation releases the stress on some processor subsystem and does not increase the stress on another one, it does not need to be autotuned because it never degrades the performance.}. For example, thread coarsening tuning mentioned in the previous section decreases the number of arithmetic operations and memory footprint (decreases the stress on floating-point units and memory subsystem), whereas it increases register consumption and decreases parallelism (increases stress on the latency-hiding mechanism and strong scaling). Another example would be employing the shared memory for a variable, which decreases the amount of global memory accesses but increases shared memory accesses (decreasing stress on L2 cache and global memory while increasing stress on shared memory).

The profiling counters (PCs) capture the workload of different hardware subsystems. Although each vendor implements their own PCs, we can distinguish two fundamentally different categories of PCs: (i) the counters measuring the stress on a processor subsystem, called $\mli{PC_{stress}}$, and (ii) the counters measuring the number of operations performed, or the amount of resources used on the subsystem, called $\mli{PC_{ops}}$. For example, $\mli{PC_{stress}}$ can contain PCs measuring the relative utilization of floating-point units or global memory bandwidth, whereas $\mli{PC_{ops}}$ can contain the number of floating-point instructions or memory instructions.

The value of a profiling counter $\mli{pc} \in \mli{PC}$ depends, in general, on all values of tuning parameters in $\mli{TP}$, the input $i \in I$ and the GPU hardware $\mli{gpu} \in \mli{GPU}$. We can formalize this relation by defining the function $f$:
\begin{equation}
f: \mli{TP} \times I \times {GPU} \rightarrow \mli{PC}
\end{equation} 
For simplicity, we assume that the compiler version or switches stay the same. The dependence of the function $f$ on $I$ and $\mli{GPU}$ differs when we distinguish between $\mli{PC_{ops}}$ and $\mli{PC_{stress}}$. The value of $\mli{PC_{stress}}$ is strongly dependent on the GPU model and input: for example, when a GPU with a higher flop-to-word ratio is used, the utilization of floating-point units can be lower as the profiled kernel becomes memory-bound. On the other hand, the amount of floating-point instructions, measured by a profiling counter of type $\mli{PC_{ops}}$, depends on the GPU model weakly (it can be affected by the GPU instruction set, but not greatly). Therefore, we can define function $g$:
\begin{equation}
g: \mli{TP} \times I \rightarrow \mli{PC}
\end{equation} 
as an approximation of $f$ independent on GPU hardware: 
\begin{equation}
\forall \mli{tp} \in \mli{TP} \ \forall i \in I \  \forall \mli{gpu} \in \mli{GPU}: g(\mli{tp}, i) \approx f(\mli{tp}, i, \mli{gpu}).
\end{equation} 
Moreover, the partial derivative of the function $g$ w.r.t. input is not changed significantly: 
\begin{equation}
\forall \mli{tp} \in \mli{TP}: \frac{\partial g}{\partial i}(\mli{tp}, i_1) \approx \frac{\partial g}{\partial i}(\mli{tp}, i_2)
\end{equation}
considering inputs $i_1$ and $i_2$ of different sizes. In other words, when some tuning parameter changes the number of operations performed on a particular processor subsystem, it does it independently on the input size. For example, thread coarsening decreases the number of floating-point operations in some ratio, although the actual amount of floating-point operations is determined by the input size. 

Approximating $f$ by $g$ suffers some imprecision in $\mli{PC_{ops}}$ related to cache and memory subsystem. Since different GPU architectures have different cache sizes, the threshold of cache capacity-misses differs across GPU architectures. Therefore, TP controlling cache blocking will start to increase the amount of operations on different cache levels at a different value of TP. However, this imprecision should only be observed within a small range of TP values, which causes the cache footprint to be close to the GPU cache capacity. This effect can be seen in our example shown in Figure~\ref{fig:dependence-ilustration}, where the amount of floating-point instructions and texture cache transactions is similar on both experiments shown in the Figure~\ref{fig:dependence-ilustration}, whereas the L2 cache workload is changed by the tuning parameter more significantly when solving a larger gridbox on GTX 1070. 

This observation on the relationship between TPs, input size, hardware model, and PCs is essential for the proposed searcher. The searcher can use a model of the stable relationship between TPs and $\mli{PC_{ops}}$ created on any GPU and input, and navigate tuning space searching according to the actually measured $\mli{PC_{stress}}$ and $\mli{PC_{ops}}$.

\subsection{Basic idea}
\label{sect:basic_idea}
Our search method aims to mimic the process of the developer when optimizing performance. Similarly to the developer, the method profiles the tuned code, analyzes observed bottlenecks and decides how to change the tuning parameters to suppress them.

We capture the developer's expertise regarding how TPs relate to PCs in a \textit{model}, and we capture how performance counters relate to bottlenecks in an \textit{expert system}. The model describes the relation between TPs and $\mli{PC_{ops}}$, which remains stable with respect to GPU hardware and input changes, so the model does not need to be re-trained. We can view the model as a component that allows the searcher to understand how changing the tuning parameters decreases or increases the stress on a processor subsystem (but cannot predict how stressed the subsystem would be: this is measured by $\mli{PC_{stress}}$ during tuning). When any tuning configuration is executed, performance counters from both $\mli{PC_{ops}}$ and $\mli{PC_{stress}}$ are measured. The expert system then identifies bottlenecks: it deduces which subsystems are overloaded with the current combination of tuning configuration, input and GPU. 

As the model knows the relation between TPs and $\mli{PC_{ops}}$, the searcher can select a tuning configuration which decreases the stress on the overloaded subsystems (by decreasing the number of operations on those subsystems measured by $\mli{PC_{ops}}$). This workflow can be seen in our example manual search, described in Section~\ref{sect:example_search} -- the developer measured the actual utilization of texture cache, floating-point units, and GPU occupancy (measured by counters from $\mli{PC_{stress}}$). Then, they changed thread coarsening in order to decrease the amount of texture memory operations, floating-point operations, or to increase parallelism (measured by counters from $\mli{PC_{ops}}$).


Summarizing, we formulate three assumptions, which can be empirically tested:
\begin{enumerate}
  \item it is possible to determine the computational bottlenecks from the values of performance counters (both $\mli{PC_{stress}}$ and $\mli{PC_{ops}}$);
  \item it is possible to determine which $\mli{PC_{ops}}$ need to be changed to suppress the bottlenecks;
  \item the relation between TP and $\mli{PC_{ops}}$ is sufficiently portable across different GPUs and problem inputs.
\end{enumerate}

The first and second assumptions are based on knowledge of how performance counters work. If something seriously hinders the performance, performance counters should be able to reflect that in some manner (the first assumption). When the bottleneck is known, it is possible to determine which $\mli{PC_{ops}}$ reflects it and therefore, which type of operations should be reduced to improve the performance. We aim to mirror this process by creating an expert system for bottleneck analysis using available documentation and our expertise -- details are in Section~\ref{sect:bottlenecks}. We evaluate these two assumptions experimentally in Section~\ref{eval:pc_biasing}.

The third assumption has been discussed in Section~\ref{sect:pcs}. We further evaluate the third assumption experimentally in Section~\ref{eval:hw_portability} and Section~\ref{eval:input_portability}.

Overall, if all of our assumptions are true, it is possible to mimic the process of a developer's performance optimization during autotuning using the following steps:
\begin{itemize}
 \item start with a certain tuning configuration;
 \item measure PCs of the tuned kernel, \eg{}, the amount of local memory used;
 \item determine the bottlenecks using the expert system, \eg{}, the bandwidth on multiple levels of the memory hierarchy is high;
 \item determine the desired change in PCs (denoted as $\Delta \mli{PC}$ from now on), with the expert system, \eg{}, lower the amount of local memory used;
 \item evaluate which tuning configurations change PCs in the desired way using the model, \eg{}, those which have lower \texttt{Z\_ITERATIONS} than the profiled configuration;
 \item select the next tuning configuration.
\end{itemize}
We replace the selection of the next tuning configuration, usually done by the developer, with a step of random search biased by a score determining the likelihood of changing PCs in the required direction.

To sum up, we move from given values of TPs (defining a tuning configuration) through PCs (performance counters) to bottlenecks, and then from bottlenecks through $\Delta \mli{PC}$ (changes to performance counters) back to the new set of TP (new tuning configuration).

\subsection{The architecture}

\begin{figure}[t]
    \centering
    \includegraphics[width=.99\hsize]{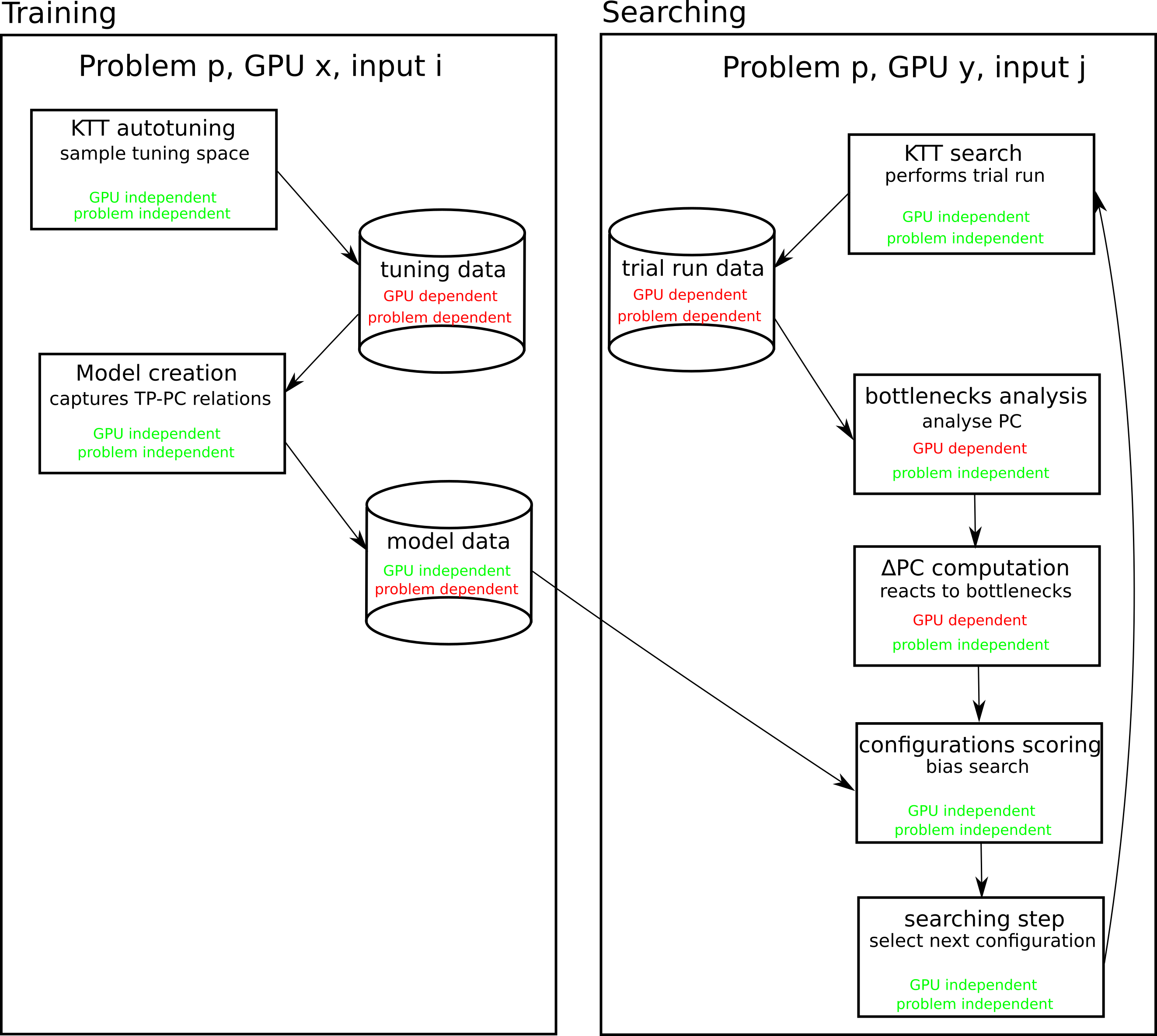}
    \caption{Schematic view of the searcher workflow. The boxes show program components, cylinders show data objects.}
    \label{fig:searcher}
\end{figure}

The architecture of the whole tuning workflow is depicted in Figure~\ref{fig:searcher}. It consists of two phases: one for training beforehand and one for the actual search during autotuning. 

During the training phase, two components are used. KTT is responsible for executing tuning and gathering PCs on a sample or a complete tuning space. Then, a model is built from the tuning data using any of the methods described in Section~\ref{sect:models}. Both components are problem- and GPU-independent. The raw tuning data are problem-, GPU- and input-dependent. The model is still problem-dependent (different problems can have completely different TP), but, according to the third hypothesis, it is independent on GPU model and input. 

During the autotuning for GPU and input of our interest, KTT is used to collect $\mli{PC_{stress}}$ and $\mli{PC_{ops}}$. Then, the component called \textit{bottleneck analysis} analyzes $\mli{PC_{stress}}$ to determine bottlenecks. It also uses $\mli{PC_{ops}}$ to distinguish the source of the bottleneck more precisely (\eg{}, if global memory bandwidth is the bottleneck, it uses the number of memory transactions to recognize how much the bottleneck is caused by memory reads or stores). The \textit{$\Delta \mli{PC}$ computation} component determines the required change of $\mli{PC_{ops}}$. The bottleneck analysis and $\Delta \mli{PC}$ computation components are GPU-dependent, because performance counters have been changed or extended several times as GPUs have evolved. Therefore, those components contain explicit support for different sets of performance counters. It is important to note that the bottleneck component analyzes the bottlenecks of the GPU architecture which is used for autotuning, whereas changes of $\mli{PC_{ops}}$ are computed for the GPU architecture the model was built on during the training phase. Therefore, we can use one model to steer autotuning on multiple architectures\footnote{An alternative implementation would be to translate $\mli{PC_{ops}}$ used in the model for the GPU architecture used with autotuning.}. When the required changes of $\mli{PC_{ops}}$ are determined, the component called \textit{configurations scoring} computes the scores of the tuning configurations. It uses the model to predict $\mli{PC_{ops}}$ according to TP and set higher scores to configurations which are predicted to change $\mli{PC_{ops}}$ in the required way. With the scores set, the \textit{searching step} component performs a step by selecting the next configuration for KTT to benchmark and profile.

\subsection{Modelling relation between tuning parameters and performance counters}
\label{sect:models}
In this section, we describe the models for how TPs relate to PCs. These models are created during the training phase (see the left part of Figure~\ref{fig:searcher}), \ie{}, before autotuning. The training phase is broken into two components. \textit{KTT autotuning} samples the entire or partial tuning space, and it stores the resulting TPs, PCs, and runtimes. Those data are then processed by the \textit{model creation} component, which uses a ML method to create a model of relations between TPs and $\mli{PC_{ops}}$. During autotuning, we use the model to evaluate which tuning configurations change PCs in the desired way (\ie{}, from $\Delta \mli{PC}$ to TPs), as described in detail in section~\ref{sect:scoring}.

\subsubsection{Least-square regression non-linear models}
We model the relation between TPs and PCs through non-linear regression models. As an input, we take the whole or a part of the tuning space that has been evaluated and profiled -- for each tuning configuration, performance counters are measured. We split the tuning space into subspaces determined by the values of binary tuning parameters (\eg{}, in a tuning space with three binary parameters, we split into $2^3 = 8$ subspaces). For each subspace, we select datapoints for training in a deliberate way -- for each non-binary parameter, we choose two or three values to keep the total number of value combinations relatively low, but to make the sampling of the subspace rather even despite constraints. The model for each profiling counter includes the main effects of each non-binary tuning parameter, their interactions (even those of a high order) and their influences of quadratic nature. We applied the least-square regression method to compute the coefficients. Therefore, as an output, we end up with several models for each profiling counter, where the values of binary tuning parameters determine the applicability of the model. The predictions themselves can be computed using the values of non-binary tuning parameters.

For example, consider adding a parameter controlling the use of constant memory for atoms in the example described in Section~\ref{sect:example}. Such a parameter is binary (one if constant memory is used and zero if it is not). Therefore, two models are created to capture relations between PCs and \texttt{Z\_ITERATIONS} for using and not using constant memory.

\subsubsection{Decision tree}
Decision tree builds regression or classification models in the form of a tree structure. It breaks down a dataset into smaller and smaller subsets while at the same time an associated decision tree is incrementally developed. The final result is a tree with decision nodes and leaf nodes. A decision node has two or more branches, each representing values for the attribute tested (TPs in our case). Leaf node represents a decision on the numerical target (values of PCs). The topmost decision node in a tree corresponds to the best predictor called the root node. Decision trees can handle both categorical and numerical data.

The core algorithm for building decision trees called ID3~\cite{quinlan1986induction} employs a top-down, greedy search through the space of possible branches with no backtracking. The ID3 algorithm can be used to construct a decision tree for regression by replacing Information Gain with Standard Deviation Reduction. The decision of making strategic splits affects a tree’s accuracy heavily. The decision criteria are different for classification and regression trees. Regression trees usually use mean squared error (MSE) to decide to split a node into two or more sub-nodes.

We generate a set of candidate trees. The randomly selected 50\% of the explored tuning space is used for training, and the rest is used for testing. We also alter parent nodes of the trees. We compute MAE (Mean Absolute Error) and RMSE (Root Mean Square Error) for those trees, and select the one with the lowest MAE -- in case of a tie, we select the one with lower RMSE). The selected tree is then used to predict PCs for unknown configurations.

\begin{table*}
\centering
\tiny
\begin{tabular}{|l|l|l|l|}
\hline
Counter (prior Volta)          	& Counter (Volta and newer) 						& Abbr. 	& Type \\ \hline
dram\_read\_transactions	& dram\_\_sectors\_read.sum 						& DRAM\_RT 	& Ops.\\
dram\_write\_transactions	& dram\_\_sectors\_write.sum 						& DRAM\_WT 	& Ops.\\
l2\_read\_transactions		& lts\_\_t\_sectors\_op\_read.sum 					& L2\_RT 	& Ops.\\
l2\_write\_transactions		& lts\_\_t\_sectors\_op\_write.sum 					& L2\_WT 	& Ops.\\
tex\_cache\_transactions	& l1tex\_\_t\_requests\_pipe\_lsu\_mem\_global\_op\_ld.sum 		& TEX\_RWT 	& Ops.\\
local\_memory\_overhead		& l1tex\_\_t\_sectors\_pipe\_lsu\_mem\_local\_op\_st.sum 		& LOC\_O 	& Ops.\\
shared\_load\_transactions	& l1tex\_\_data\_pipe\_lsu\_wavefronts\_mem\_shared\_op\_ld.sum 	& SHR\_LT 	& Ops.\\
shared\_store\_transactions	& l1tex\_\_data\_pipe\_lsu\_wavefronts\_mem\_shared\_op\_st.sum 	& SHR\_WT 	& Ops.\\
inst\_fp\_32			& smsp\_\_sass\_thread\_inst\_executed\_op\_fp32\_pred\_on.sum 		& INST\_F32 	& Ops.\\
inst\_fp\_64			& smsp\_\_sass\_thread\_inst\_executed\_op\_fp64\_pred\_on.sum		& INST\_F64 	& Ops.\\
inst\_integer			& smsp\_\_sass\_thread\_inst\_executed\_op\_integer\_pred\_on.sum 	& INST\_INT 	& Ops.\\
inst\_misc			& smsp\_\_sass\_thread\_inst\_executed\_op\_misc\_pred\_on.sum		& INST\_MISC 	& Ops.\\
inst\_compute\_ld\_st		& smsp\_\_sass\_thread\_inst\_executed\_op\_memory\_pred\_on.sum	& INST\_LDST 	& Ops.\\
inst\_control			& smsp\_\_sass\_thread\_inst\_executed\_op\_control\_pred\_on.sum	& INST\_CONT 	& Ops.\\
inst\_bit\_convert		& smsp\_\_sass\_thread\_inst\_executed\_op\_conversion\_pred\_on.sum	& INST\_BCONV 	& Ops.\\
inst\_executed			& smsp\_\_inst\_executed.sum						& INST\_EXE	& Ops.\\
issue\_slot\_utilization	& smsp\_\_issue\_active.avg.pct\_of\_peak\_sustained\_active 		& INST\_ISSUE\_U	& Ops.\\
dram\_utilization		& dram\_\_throughput.avg.pct\_of\_peak\_sustained\_elapsed $: 10$ 	& DRAM\_U 	& Stress\\
l2\_utilization			& lts\_\_t\_sectors.avg.pct\_of\_peak\_sustained\_elapsed 		& L2\_U 	& Stress \\
tex\_utilization		& l1tex\_\_t\_requests\_pipe\_lsu\_mem\_global\_op\_ld.avg.pct\_of\_peak\_sustained\_active $:10$ & TEX\_U 	& Stress \\
shared\_utilization		& l1tex\_\_data\_pipe\_lsu\_wavefronts\_mem\_shared.avg.pct\_of\_peak\_sustained\_elapsed $:10$ & SHR\_U  	& Stress\\
sm\_efficiency			& smsp\_\_cycles\_active.avg.pct\_of\_peak\_sustained\_elapsed		& SM\_E 	& Stress\\
warp\_execution\_efficiency	& smsp\_\_thread\_inst\_executed\_per\_inst\_executed.ratio $\cdot 100 : 32$	& WARP\_E	& Stress\\
warp\_nonpred\_execution\_efficiency 	& smsp\_\_thread\_inst\_executed\_per\_inst\_executed.pct 	& WARP\_NP\_E 	& Stress\\
\hline
\end{tabular}
\caption{List of performance couters and their abbreviations for GPUs. For counters implemented for Volta generation and newer, the conversion ratio (if any) is written next to the counter.}
\label{tab:pcs}
\end{table*}

\subsection{Bottleneck detection and reaction}
\label{sect:bottlenecks}
In this section, we describe how we get from PCs to bottlenecks and from bottlenecks to $\Delta \mli{PC}$. This part is performed by expert systems which read performance counters, analytically determine bottlenecks, and create a vector of required changes of $\mli{PC_{ops}}$. 

The components described in this section do two steps -- the first component takes performance counters measured on an actually executed kernel and computes bottlenecks. The second component takes bottlenecks and  computes changes in performance counters for the architecture of the model. Therefore, the components have to know the PCs of the currently executed kernel, the architecture of the GPU where the kernel was executed, and the architecture of the GPU the model was created for.

Since the performance counters changed completely for Volta generation and newer, the components are implemented for multiple sets of counters. The implementation is limited by our understanding of NVIDIA GPU's performance counters, and we believe there is still room for improvement if the counters were documented in greater detail, or if the entire expert system was replaced by a machine learning model. We have developed the expert system using NVIDIA CUPTI documentation for CUDA 10.0. To improve our understanding of the counters' values, we have also used some benchmarks available in KTT. However, to avoid overfitting the expert system to our benchmarks, we have only used three benchmarks for its development (Coulomb sum, n-body and Matrix transposition), while the rest of the benchmarks have been used for evaluation of our searcher after it was developed.

\subsubsection{Bottleneck analysis}

The bottleneck analysis uses a set of performance counters described in Table~\ref{tab:pcs}. We use abbreviations of those counters in the following text and formulas. The table also shows the equivalency of old (\ie{}, on GPUs prior to Volta) and new (\ie{}, on Volta and newer) counters; applied adjustments or rescalings (\eg{}, the old version of counters measures rank in $<0,10>$, whereas the new version measures percentage in $<0, 100>$); and classification of counters into $\mli{PC_{stress}}$ or $\mli{PC_{ops}}$. Note that the equivalence of new and old performance counters is not completely documented, so we compute it according to our understanding of what counters measure. Also note that we have assigned INST\_ISSUE\_U into $\mli{PC_{ops}}$, because it quantifies the ratio of instruction cycles which cannot be used to issue instructions (for instance, due to synchronizations). 

Bottlenecks are represented by a vector $B = [\mli{b_{DRAM\_read}},\\ \mli{b_{DRAM\_write}}, \cdots ]$, where $\forall b_x \in B, b_x \in <0, 1>$. A zero value of the bottleneck means that the component is not stressed, while the value 1 represents the component being on the theoretical peak of its performance. Bottlenecks are divided into several categories, representing stress on various memory types, instructions and utilization of GPU parallelism.

The memory subsystems bottleneck analysis is computed in the same way for global memory, shared memory and L2 cache. The utilization of the memory, reported by performance counter DRAM\_U, is scaled into $<0, 1>$ and weighted by the ratio of read and write transactions. Equation~\ref{eq:global-read} shows the computation of the bottleneck for global memory read, Equation~\ref{eq:global-write} shows the computation of the bottleneck for global memory write. For shared memory and L2 cache, the computation is analogous, using SHR\_LT, SHR\_WT, SHR\_U and L2\_RT, L2\_WT, L2\_U, respectively.

\begin{equation}
\footnotesize
\mli{b_{DRAM\_read}} = \frac{\mli{DRAM\_RT}}{\mli{DRAM\_RT} + \mli{DRAM\_WT}} \cdot \frac{\mli{DRAM\_U}}{10}
\label{eq:global-read}
\end{equation}

\begin{equation}
\footnotesize
\mli{b_{DRAM\_write}} = \frac{\mli{DRAM\_WT}}{\mli{DRAM\_RT} + \mli{DRAM\_WT}} \cdot \frac{\mli{DRAM\_U}}{10}
\label{eq:global-write}
\end{equation}

To compute the utilization of texture (data) cache, we only scale the counter TEX\_U to interval $<0, 1>$, as the cache is read-only and therefore there are no write transactions. 

The situation is more complicated with local memory. To the best of our knowledge, the LOC\_O performance counter gives us a relative number of local memory data transfers. Therefore, even high local memory overhead does not imply the local memory is a real bottleneck if no memory subsystem is overloaded. Consequently, we weight local memory overhead by the maximal utilization of memories in the way: global, L2 and texture, see Equation~\ref{eq:local}.

\begin{equation}
\footnotesize
\mli{b_{local}} = \frac{\mli{LOC\_O}}{100} \cdot \max(\frac{\mli{DRAM\_U}}{10}, \frac{\mli{L2\_U}}{10}, \frac{\mli{TEX\_U}}{10})
\label{eq:local}
\end{equation}

Special performance counters measure the utilization of instructions. However, we found those counters not to be very reliable, or we do not fully understand what they measure (for example, the matrix transposition example, which uses no floating point computations, reports high utilization of FP32 units in some configurations). Therefore, we derive the utilization from the overall amount of instructions. First, we compute the $\mli{ins_{fitted}}$ value using Equation~\ref{eq:inst_fitted}, which computes the number of instructions (INST\_EXE computes at the warp level; therefore, it has to be multiplied by 32), corrected by the efficiency of instruction execution (\ie{}, how many threads in the warp perform useful work). 

\begin{equation}
\footnotesize
\mli{ins_{fitted}} = 32 \cdot \mli{INST\_EXE} \cdot \frac{100}{\mli{WARP\_E}} \cdot \frac{100}{\mli{WARP\_NP\_E}}
\label{eq:inst_fitted}
\end{equation}

Then, we compute the utilization of the issue slot. For GPUs prior to Volta, we use $\mli{ins_{util}} = \frac{\mli{INST\_ISSUE\_U}}{100}$. For Volta and newer, as those GPUs allow us to issue integer and floating-point instructions separately, we compute $\mli{ins_{util}} = \min(1, \frac{\mli{INST\_ISSUE\_U}}{50})$. Therefore, the component considers full utilization of one instruction path as perfect utilization and cannot optimize toward better utilization of dual-issue in the current implementation. The utilization of FP32 units is computed using Equation~\ref{eq:fp32}. An analogous computation is performed for INST\_F64, INST\_INT, INST\_MISC, INST\_LDST, INST\_CONT and INST\_BCONV.

\begin{equation}
\footnotesize
\mli{b_{fp32}} = \frac{\mli{INST\_F32}}{\mli{ins_{fitted}}} \cdot \mli{ins_{util}}
\label{eq:fp32}
\end{equation}

Finally, the bottleneck formed by the insufficient instruction issue utilization is computed. First, we compute maximal utilization across all types of instructions:
\begin{equation}
\footnotesize
\mli{util_{max}} = \max(\frac{\mli{INST\_F32}}{\mli{ins_{fitted}}}, \frac{\mli{INST\_F64}}{\mli{ins_{fitted}}}, \dots)
\label{eq:maxutil}
\end{equation}
Then, the bottleneck of instruction issue is computed using Equation~\ref{eq:issue}.
\begin{equation}
\footnotesize
\mli{b_{issue}} = \mli{util_{max}} \cdot \frac{100 - \mli{INST\_ISSUE\_U}}{100}
\label{eq:issue}
\end{equation}

Insufficient parallelism is computed as 
\begin{equation}
\footnotesize
\mli{b_{sm}} = \frac{100 - \mli{SM\_E}}{100}.
\end{equation} 

An additional bottleneck of insufficient parallelism is computed from the number of CUDA threads and the number of CUDA cores, as shown in Equation~\ref{eq:paral}. It is an empirical computation, requiring five threads per CUDA core to set the parallelism bottleneck to zero. However, we have found it useful to add this bottleneck alongside $\mli{b_{sm}}$, as it captures the cases where all SMs are occupied, but the actual occupancy is low due to small number of threads running per SM. 
\begin{equation}
\footnotesize
\mli{b_{paral}} = max(0, (\mli{cores} \cdot 5 - \mli{threads}) / (\mli{cores} \cdot 5))
\label{eq:paral}
\end{equation}

\subsubsection{$\Delta \mli{PC}$ computation}
\label{sect:reaction}

When the bottleneck vector is computed, it is passed to the component responsible for computing required changes to PCs. The required changes are represented as the vector $\mli{\Delta PC_{ops}} = [\mli{\Delta pc_{DRAM\_RT}},\- \mli{\Delta pc_{DRAM\_WT}},\- \dots]$, where $\forall \mli{\Delta pc_x} \in \mli{\Delta PC_{ops}}, \mli{\Delta pc_x} \in <-1, 1>$. The $\mli{\Delta PC_{ops}}$ vector contains the required changes to the performance counters $\mli{PC_{ops}}$. A negative value means that the counter value should be decreased, while a positive value means that the counter should be increased -- zero means no change is required.

There is one additional parameter for $\mli{\Delta PC_{ops}}$ computation: $\mli{inst\_reaction}$. It determines the threshold of in\-struc\-tion-re\-lat\-ed bottleneck values, which triggers the changes of $\mli{PC_{ops}}$. The motivation behind the introduction of $\mli{inst\_re}\-\mli{action}$ parameter is as follows. The instructions have very low latency (compared to the memory subsystem). Therefore, they do not form a serious bottleneck unless there is high stress on the component performing the instructions. In case of no significant bottleneck for the kernel, the low bottlenecks related to instructions are ignored, and the memory is optimized, which should decrease latency and introduce a new bottleneck, either instruction- or memory-related. In our implementation, the $\mli{inst\_reaction}$ is set to 0.7. If user sets that the instruction-bound problem is optimized, $\mli{inst\_reaction}$ is set to 0.5, which slightly improves results for compute-bound problems as the reaction on appearing instructions-related bottleneck is faster.

The changes of PCs related to the memory subsystems are computed straightforwardly: their value is set as the inverse value of the corresponding bottleneck, for example, $\mli{\Delta pc_{DRAM\_RT}} = - \mli{b_{DRAM\_read}}$. 

The instruction-related PCs are set only if the corresponding bottleneck value exceeds $inst\_reaction$, as shown in Equation~\ref{eq:fp32_reaction}. An analogous computation is performed for all instruction-related bottlenecks including $\mli{b_{issue}}$.
\begin{equation}
\begin{split}
\footnotesize
\mli{b_{fp32}} \leq \mli{inst\_reaction} &:	\scriptstyle \mli{\Delta pc_{INST\_F32}} = 0 \\
\footnotesize
\mli{b_{fp32}} > \mli{inst\_reaction} &: \scriptstyle \mli{\Delta pc_{INST\_F32}} = - \frac{(\mli{b_{fp32}} - \mli{inst\_reaction})}{1 - \mli{inst\_reaction}} 
\label{eq:fp32_reaction}
\end{split}
\end{equation}

The parallelism-related bottlenecks are applied straightforwardly, without inverting the bottleneck value: $\mli{\Delta pc_{SM\_E}} = \mli{b_{sm}}$ and $\mli{\Delta pc_{global}} = \mli{b_{paral}}$. Note that $\mli{\Delta pc_{global}}$ is not a true hardware performance counter, but represents the number of threads reported by KTT, which is added to the set of performance counters.

\subsection{Configurations scoring}
\label{sect:scoring}

Let $\mli{c_{profile}}$ be the configuration which has been empirically profiled during autotuning. Let $\mli{\Delta PC}$ be the required changes of PCs to soften bottlenecks (see Section~\ref{sect:reaction}) of the configuration $\mli{c_{profile}}$. To score an unexplored tuning configuration $\mli{c_{candidate}}$, we need to estimate whether the configuration $\mli{c_{candidate}}$ changes PCs (compared to $\mli{c_{profile}}$) in a direction defined by $\mli{\Delta PC_{ops}}$. 

Since autotuning can be executed on a different GPU and different input, we cannot directly compare estimated PCs of $\mli{c_{candidate}}$ to measured PCs of $\mli{c_{profile}}$. Instead, we use a model described in Section~\ref{sect:models} to estimate PCs of both configurations: let $\mli{PC}(\mli{c_{profile}})$ be PCs of $\mli{c_{profile}}$ estimated by the model and $\mli{PC}(\mli{c_{candidate}})$ be PCs of $\mli{c_{candidate}}$ estimated by the model. The score $s$ of the configuration $\mli{c_{candidate}}$ is computed using Equation~\ref{eq:score}. 

\begin{equation}
\footnotesize
s = \sum_{p \in \mli{PC_{used}}} \mli{\Delta pc}_p \frac{\mli{pc}_p(\mli{c_{profile}}) - \mli{pc}_p(\mli{c_{candidate}})}{\mli{pc}_p(\mli{c_{profile}}) + \mli{pc}_p(\mli{c_{candidate}})}
\label{eq:score}
\end{equation}
where $\mli{PC_{used}}$ is the set of profiling counters with non-zero predictions for profiled and candidate configurations: $\mli{PC_{used}} = \forall p \in \mli{\Delta PC} : \mli{pc}_p(\mli{c_{profile}}) \neq 0 \wedge \mli{pc}_p(\mli{c_{candidate}}) \neq 0$ and where $\mli{pc_p}$ means performance counter $p$.

We compute all scores $s_0 \dots s_n$ for configurations of our interest. Then, the score values are further processed. The lowest and highest score are stored in $s_{min}$ and $s_{max}$, respectively. The score for each configuration is normalized into interval $<0.0001, 256>$ as follows:
\begin{equation}
\begin{split}
\footnotesize
s > 0 &: \scriptstyle \mli{s_{norm}} = (1 + \frac{s}{\mli{s_{max}}})^8 \\
\footnotesize
s > \gamma \wedge s \leq 0 &: \scriptstyle \mli{s_{norm}} = \max(0.0001, (1 - \frac{s}{\mli{s_{min}}})^8) \\
\footnotesize
s \leq \gamma &: \scriptstyle \mli{s_{norm}} = 0.0001 \\
\label{eq:score_normalization}
\end{split}
\end{equation}

The normalized score is used to bias the probability of selecting that configuration. Equation~\ref{eq:score_normalization} ensures strong preference of high scores (positive scores are amplified into $<1, 256>$ by normalization). However, it also allows for the selection of negative scores, albeit with very low probability. If the score before normalization is below cutoff threshold $\gamma$ (defined as $-0.25$ in our implementation), the normalized score is set to 0.0001. Negative scores greater than $\gamma$ are scaled similarly to the positive scores. Setting non-zero probability for negative scores is important in situations where all unexplored configurations seem worse than $\mli{c_{profile}}$, which either indicates that the tuning has reached the optimum, or that it got stuck in a local optimum due to inaccuracy in the model or in the expert system decision.

\subsection{Searching step}



Our current implementation uses weighted random search, where the random selection is biased towards configurations which should soften bottlenecks found within the last profiled configuration. It is the most straightforward approach, which allows for direct comparison to (unweighted) random search. However, we believe that more sophisticated searching methods could be implemented in the future.

The complete workflow of the search is described by Algorithm~\ref{alg:tuning}. The algorithm takes several variables as input: $\mli{TS}$ is a tuning space (\ie{}, an array containing all tuning configurations), $M$ is a model of the relation between TPs and PCs, $n$ is the number of steps performed without collecting performance counters and $i$ is the number of iterations for collecting performance counters (\ie{}, the number of empirical tests is $i(n + 1)$). Note that CUDA kernels are executed faster when no performance counters are collected. Therefore, executing runs without collecting performance counters helps to find a fast kernel more quickly, although with the cost of performing more empirical search steps. The default value of $n$ is set to 5 in our implementation. 

Algorithm~\ref{alg:tuning} uses the following components. In Line~\ref{ln:ktt}, the empirical evaluation of kernel $\mli{c_{profile}}$ is performed and its runtime and PCs are stored into $t$, $\mli{PC_{stress}}$ and $\mli{PC_{ops}}$. Then, bottlenecks are determined and stored in $b$ in line~\ref{ln:bottlenecks}. After bottleneck detection, the searcher computes a reaction to the bottlenecks in line~\ref{ln:changes}. The reaction is stored in the form of the required change of performance counters $\mli{\Delta PC_{ops}}$. In the current implementation, the algorithm scores all tuning configurations according to the required change of performance counters: it uses model $M$ to determine if a tuning configuration changes performance counters in a direction required by $\mli{\Delta PC_{ops}}$, see line~\ref{ln:model}~\footnote{In the case of huge tuning spaces, a restricted set of tuning configurations could be scored, \eg{}, only the neighbourhood of the currently explored configuration $\mli{c_{profile}}$.}. With the scored tuning space, the algorithm selects $n$ kernels to benchmark without gathering performance counters. In the loop starting at line~\ref{ln:nonprofile}, the algorithm selects a configuration with a probability weighted by the score obtained in line~\ref{ln:model}. Then, the benchmarking of the selected configuration is performed. The fastest observed configuration is used for the next profiling run (performed in line~\ref{ln:ktt}).

\begin{algorithm}
  \KwIn{$TS, M, n, i$}
  $\mli{c_{profile}} \leftarrow$ random $c \in \mli{TS}$ \\
  \For{$j = 0; j \le i; j++$} {
    $t, \mli{PC_{stress}}, \mli{PC_{ops}} \leftarrow$ empiricaly measure runtime and PCs of $\mli{c_{profile}}$ \label{ln:ktt} \\
    $b \leftarrow$ bottlenecks obtained from $\mli{PC_{stress}}$ and $\mli{PC_{ops}}$\label{ln:bottlenecks} \\
    $\mli{\Delta PC_{ops}} \leftarrow $ changes in PCs required to weaken bottlenecks $b$ \label{ln:changes} \\
    $S \leftarrow {0}^{\mli{size}(\mli{TS})}$ \\
    \For{$k = 0; k < \mli{size}(\mli{TS}); k++$} {
      \If {$\mli{TS_k}$ was not evaluated} {
        $S_k \leftarrow$ compute score of $\mli{TS_k}$, using $M$ and $\mli{\Delta PC_{ops}}$ \label{ln:model} \\
      }
      \Else {
        $S_k \leftarrow 0$
      }
    }
    $t \leftarrow \infty$\\
    \For{$k = 0; k \le n; k++$} { \label{ln:nonprofile}
      $r \leftarrow$ random number: $r \leq 0 \wedge r < \sum_{l=0}^{\mli{size}(\mli{TS})}S_l$ \\
      select $l$ such that $\sum_{i=0}^{l-1}S_i \geq r \wedge \sum_{i=0}^{l}S_i < r)$ \\
      $t' \leftarrow$ empirically measure runtime of $\mli{TS_l}$ \\
      \If {$t' \leq t$} {
        $\mli{c_{profile}} \leftarrow \mli{TS_l}$ \\
        $t \leftarrow t'$\\
      }
      $S_l \leftarrow 0$ \\
    }
  }
  \caption{Searching tuning space with performance counters.}
  \label{alg:tuning}
\end{algorithm}

\subsection{Implementation}
The main part of the proposed profile-based searcher is implemented in Python in its current version. As KTT is implemented in C++, there is a simple stub searcher code in KTT, which communicates with the main part of the searcher via files and sockets. The reason for Python implementation is easier experimenting with the searcher code as well as a broad range of easy-to-use machine learning algorithms. Nevertheless, such implementation has also a disadvantage in code performance. It is possible to rewrite the searcher code into C++ in the future.

The searcher consists of two parts: an application building the model and a plugin into KTT, which implements the proposed searcher method. The searcher is available in \texttt{profile-searcher} directory in KTT\footnote{GIT tag \texttt{v1.3-profile-searcher}}.

\subsection{Limitations of the current implementation}
\label{sect:limits}

Although the proposed profile-based searcher is fully capable of searching the tuning space, biasing the search towards faster implementations, there are still many open research topics which can lead to the improvement of the searcher.

\subsubsection{Using local search method}
Currently, our implementation biases a simple random searcher. Although a random searcher is very robust, it is a global search method, which cannot speed up its convergence by following the gradient of the optimized function. Our searcher allows us to estimate which configurations should improve performance. This estimation could be used as an estimation of the gradient of the performance function, and thus a gradient-based searching method could be used. As shown in~\cite{vanwerkhoven2018kernel}, local searching methods can be successful, especially in combination with a global searching method, which allows them to overcome local optima.

\subsubsection{Improving model of TPs-PCs relations}
Currently, the relations of TPs to PCs are always modelled for the GPU where the model was constructed. Although it is sufficient to steer the tuning space search, it cannot precisely model the effect of cache capacity misses. We believe that it is possible to create a model which can correct cache-related PCs according to detected caching capabilities of the GPU used for autotuning.

\subsubsection{Improving bottleneck analysis and reaction}
Our current expert system for bottleneck analysis and computation of the required changes of PCs works with a subset of PCs which are available on GPUs. For example, PCs related to execution stall reasons are not used at all. Although it is not easy to construct an expert system which would interpret all PCs, it could be possible to replace the expert system with a machine learning model which could overcome our limited understanding of PCs interpretation. To construct the model, we can use multiple benchmarks and GPUs available. Moreover, each benchmark consists of a high number of tuning configurations; therefore, it is not too difficult to generate a high volume of training and testing data.

The runtime of kernel profiling is determined by the type and amount of gathered hardware performance counters. Together with improving bottleneck analysis and reaction system, we could test if some counters are mostly redundant and reduce their amount, similarly to~\cite{alcaraz2019hardware}.

\section{Evaluation}
\label{sect:eval}
In this section, we evaluate our proposed profile-based searcher. First, we describe the methodology of the benchmarks and the testbed setup. Then, we evaluate how random step biasing influences the searcher performance when using a model created for the same setup, for a different GPU or for a different input. We also evaluate the convergence time of the proposed method. Finally, we compare our method with model-based tuning using Starchart~\cite{jia2013starchart} and with optimization-based tuning using Kernel Tuner~\cite{vanwerkhoven2018kernel}.

\subsection{Methodology}
We have performed two types of evaluation:
\begin{itemize}
  \item measuring the number of empirical tests of kernels (\ie{}, searcher steps);
  \item measuring tuning convergence in time.
\end{itemize}

The empirical test means that the kernel is compiled and executed in the selected tuning configuration. The number of empirical tests allows us to directly compare the efficiency of the proposed profile-based search method: how many empirical tests it needs to perform in order to converge to a certain efficiency of the kernel. On the other hand, the GPU kernel runs slower when being profiled as it gathers PCs. Therefore, even reducing the number of empirical tests does not ensure faster tuning convergence of the proposed searcher. Therefore, we have also tested tuning convergence speed. In our evaluations, we often aim to find a well-performing configuration. We define a well-performing configuration as a configuration which creates a kernel with runtime within $1.1\times$ of the best kernel runtime (found by exhaustive search).

Since the tuning space search is a stochastic process, it has to be repeated many times to get results which are free from random fluctuations. As it would be incredibly demanding to actually run and profile kernels during autotuning, we have created a program bypassing that -- instead of running kernels many times, it performs an exhaustive exploration of the entire tuning space and saves the tuning results (kernel runtimes and PCs). Then we can perform autotuning space search faster, \ie{} simply load the kernel runtimes and PCs from files and provide them to the searcher instead of actually running the kernel and profiling it. This allows us to repeat tuning $1000\times$ for any combination of benchmark, hardware used to create the model, and hardware used for autotuning. We only used this simulated autotuning to evaluate the number of searcher steps. 

While we were evaluating tuning convergence time, we always performed empirical testing, \ie{}, we actually ran and profiled the kernels and measured the time. Therefore, we executed those tests $100\times$ due to much higher time demands.

\subsection{Testbed Setup}

\begin{table}
\centering
\footnotesize
\begin{tabular}{|l|r|r|}
    \hline
    Benchmark 	& dimensions & configurations \\
    \hline
    Convolution	& 10 & 3,928 \\
    Coulomb 3D  & 7 & 210 \\
    GEMM	& 10 & 5,788 \\
    GEMM full   & 14 & 205,216 \\
    Transpose   & 8 & 1,784 \\
    N-body	& 7 & 3,134 \\
    \hline
\end{tabular}
\caption{A list of the benchmarks and the size and dimensionality (\ie{}, the number of tuning parameters) of their tuning spaces.}
\label{tab:benchmarks-spaces}
\end{table}

To test the searcher proposed in this paper, we used benchmarks listed in Table~\ref{tab:benchmarks-spaces}. Those benchmarks have been introduced in~\cite{petrovic2020benchmark}, where more details about their implementation and optimizations are discussed. We have rewritten these benchmarks into CUDA to allow their profiling on NVIDIA GPUs. We removed the values of some tuning parameters as they are not supported on CUDA (large built-in vector variables, explicit cache prefetching), or are not implemented in CUDA KTT backend (working with constant memory). However, our benchmarks are still able to reach near-peak performance on GPUs, similarly to~\cite{petrovic2020benchmark}\footnote{The highest performance degradation compared to the richer OpenCL tuning spaces was 5\% in Coulomb sum benchmark running on GeForce GTX 680. At most, 3\% performance degradation was observed in the remaining combinations of benchmarks and GPUs.}. Therefore, we consider their tuning spaces sufficient for evaluation of the tuning space searcher. We have also implemented two versions of tuning space for GEMM -- the reduced space is taken from CLBlast~\cite{nugteren2018clblast} (denoted as GEMM) and the full space is taken from CLTune~\cite{nugteren2015cltune} (denoted as GEMM full). This smaller size of GEMM space allows us to explore the entire tuning space in reasonable time, thus it is more practical for most of the experiments. 

Note that expert programmers have designed the tuning spaces to be reasonably small (without obviously slow configurations, \eg{}, resulting in sub-warp block sizes)~\cite{petrovic2020benchmark}. Such search spaces should not include a vast amount of poorly-performing tuning configurations and therefore should not discriminate simple search methods, such as random search.


\begin{table}[ht]
\centering
\footnotesize
\begin{tabular}{|l|l|l|}
\hline
Device                  & Architecture  & Released \\ \hline
GeForce GTX 680  	& Kepler        & 2012    \\
GeForce GTX 750  	& Maxwell       & 2014    \\
GeForce GTX 1070 	& Pascal        & 2016    \\
GeForce RTX 2080 	& Turing      	& 2018   \\
\hline
\end{tabular}
\caption{GPU devices used in our benchmarks.}
\label{tab:hw}
\end{table}

The benchmarks have been executed using four GPUs of different architectures, listed in Table~\ref{tab:hw}. All benchmarks have been executed with Kernel Tuning Toolkit 1.3 equipped with the profile-based searcher\footnote{\url{https://github.com/HiPerCoRe/KTT/tree/v1.3-profile-searcher}}.

\subsection{Speedup allowed by the performance counters biasing}
\label{eval:pc_biasing}

In the first experiment, we have measured the improvement given by biasing random search according to the required changes of PCs, computed by bottleneck analysis and reaction subsystems. In other words, we experimentally evaluated the first two assumptions mentioned in Section~\ref{sect:basic_idea}. To do so, we needed to eliminate any effect of the model imprecision. Therefore, we have performed exhaustive exploration of PCs of the benchmarks' complete tuning spaces and then, during autotuning, we read the previously measured PCs of tuning configurations instead of predicting them with the model. 

\begin{table}[ht]
\centering
\footnotesize
\begin{tabular}{|l|l|l|l|l|}
\hline
	        & GTX 680	& GTX 750	& GTX 1070	& RTX 2080 \\ \hline
Coulomb sum  	& $19$  	& $21$		& $34$		& $16$ \\
Matrix trans.  	& $192$  	& $24$		& $10$		& $47$ \\
GEMM	  	& $146$  	& $248$		& $450$		& $260$ \\
n-body  	& $27$  	& $10$		& $37$		& $39$ \\
Convolution  	& $327$  	& $702$		& $349$		& $568$ \\
\hline
\end{tabular}
\caption{Average number of empirical tests required for random search to find a well-performing configuration.}
\label{tab:random_iters}
\end{table}

\begin{table}[ht]
\centering
\footnotesize
\begin{tabular}{|l|l|l|l|l|}
\hline
	        & GTX 680	& GTX 750	& GTX 1070	& RTX 2080 \\ \hline
Coulomb sum  	& $3.8\times$  	& $5.25\times$	& $5.67\times$	& $3.2\times$ \\
Matrix trans.  	& $3.62\times$  & $2.0\times$	& $1.43\times$	& $1.12\times$ \\
GEMM	  	& $5.41\times$  & $7.75\times$	& $8.88\times$	& $10.83\times$ \\
n-body  	& $1.93\times$ 	& $2.5\times$	& $2.85\times$	& $3.25\times$ \\
Convolution  	& $8.18\times$  & $10.32\times$	& $15.86\times$	& $14.56\times$ \\
\hline
\end{tabular}
\caption{Improvement (in terms of the number of empirical tests) of proposed searcher over random search. The PCs from the same architecture are used.}
\label{tab:pc_bias}
\end{table}

The average number of empirical tests required by random search to find a well-performing configuration is shown in Table~\ref{tab:random_iters}. The resulting improvement of the proposed searcher over random search is shown in Table~\ref{tab:pc_bias}. The table shows improvement in terms of the average number of empirical search steps required to find a well-performing configuration. The average is obtained from 1,000 runs of the searcher.

As we can see in Table~\ref{tab:pc_bias}, the proposed searcher improves the number of empirical tests in all cases. The most significant improvement can be seen with the GEMM and Convolution benchmarks. Their tuning spaces are more complicated to search (see Table~\ref{tab:random_iters}), so the improvement given by the biased random selection is more visible. 

\subsection{Portability across hardware}
\label{eval:hw_portability}

In this section, we have evaluated the portability of the model. We created the model on specific hardware and input and used it to bias tuning space search on different hardware\footnote{In this case, we could not read PCs from a different GPU or input directly, as the executable configurations generally differ for different GPUs or inputs. Therefore, we have also had to use the predictive model.}. Therefore, we have experimentally tested the third assumption given in Section~\ref{sect:basic_idea}. The model was created using decision trees described in Section~\ref{sect:models}. Similarly to the experiment above, we have measured the improvement over the random search in terms of the number of iterations (\ie{}, empirical tests) required to reach a well-performing configuration. The results were averaged over 1,000 executions.

\begin{table}[ht]
\centering
\footnotesize
\caption{Improvement (in terms of the number of empirical tests) of proposed searcher over random search. Rows show GPUs used for benchmark execution, columns show GPUs used to create the model of TPs and PCs relations.}
\subcaption*{Coulomb sum benchmark}
\begin{tabular}{|l|l|l|l|l|}
\hline
	        & GTX 680	& GTX 750	& GTX 1070	& RTX 2080 \\ \hline
GTX 680  	& $3.8\times$ 	& $4.75\times$	& $3.8\times$	& $3.8\times$ \\
GTX 750  	& $4.2\times$   & $5.25\times$	& $5.25\times$	& $4.2\times$ \\
GTX 1070  	& $4.86\times$ 	& $5.67\times$	& $4.86\times$	& $4.86\times$ \\
RTX 2080  	& $4.0\times$   & $4.0\times$	& $4.0\times$	& $3.2\times$ \\
\hline
\end{tabular}
\newline
\subcaption*{Matrix transposition benchmark}
\begin{tabular}{|l|l|l|l|l|}
\hline
	        & GTX 680	& GTX 750	& GTX 1070	& RTX 2080 \\ \hline
GTX 680  	& $3.76\times$ 	& $5.05\times$	& $4.36\times$	& $2.95\times$ \\
GTX 750  	& $1.85\times$  & $2.0\times$	& $1.85\times$	& $0.86\times$ \\
GTX 1070  	& $1.67\times$ 	& $1.43\times$	& $1.43\times$	& $1.11\times$ \\
RTX 2080  	& $2.14\times$  & $2.76\times$	& $2.47\times$	& $1.34\times$ \\
\hline
\end{tabular}
\newline
\subcaption*{GEMM benchmark}
\begin{tabular}{|l|l|l|l|l|}
\hline
	        & GTX 680	& GTX 750	& GTX 1070	& RTX 2080 \\ \hline
GTX 680  	& $5.21\times$ 	& $2.39\times$	& $2.35\times$	& $0.26\times$ \\
GTX 750  	& $3.94\times$  & $7.52\times$	& $7.75\times$	& $8.86\times$ \\
GTX 1070  	& $4.12\times$  & $8.61\times$	& $8.61\times$	& $9.47\times$ \\
RTX 2080  	& $4.19\times$  & $18.57\times$	& $15.29\times$	& $9.63\times$ \\
\hline
\end{tabular}
\newline
\subcaption*{n-body benchmark}
\begin{tabular}{|l|l|l|l|l|}
\hline
	        & GTX 680	& GTX 750	& GTX 1070	& RTX 2080 \\ \hline
GTX 680  	& $1.93\times$ 	& $0.9\times$	& $1.17\times$	& $0.27\times$ \\
GTX 750  	& $2.0\times$   & $2.5\times$	& $2.5\times$	& $2.0\times$ \\
GTX 1070  	& $3.08\times$  & $2.64\times$	& $2.85\times$	& $1.85\times$ \\
RTX 2080  	& $3.9\times$   & $4.33\times$	& $5.57\times$	& $3.25\times$ \\
\hline
\end{tabular}
\newline
\subcaption*{Convolution benchmark}
\begin{tabular}{|l|l|l|l|l|}
\hline
	        & GTX 680	& GTX 750	& GTX 1070	& RTX 2080 \\ \hline
GTX 680  	& $9.91\times$	& $8.61\times$	& $7.79\times$	& $12.11\times$ \\
GTX 750  	& $14.93\times$ & $14.33\times$	& $11.7\times$	& $18.0\times$ \\
GTX 1070  	& $0.63\times$  & $12.46\times$	& $16.61\times$	& $7.76\times$ \\
RTX 2080  	& $14.56\times$ & $19.59\times$	& $21.04\times$	& $11.83\times$ \\
\hline
\end{tabular}
\label{mutitab:hw_port}
\end{table}

The performance portability results of Coulomb sum, Matrix transposition, GEMM, n-body and Convolution benchmarks are given in Table~\ref{mutitab:hw_port}. 
The table shows the speedup of the proposed searcher for all combinations of a GPU used to build the model and a GPU used for autotuning. 

It can be seen that the model portability is good in most of the examples. In some cases, using a model built on a different GPU is even better than using the model created on the same GPU that was used for autotuning (\eg{}, GEMM benchmark executed on RTX 2080 converges faster with the model created from GTX 1070 data). However, we consider this to be a somewhat random artefact where inaccuracies in the model compensate for inaccuracies in the expert system biasing the search. 

Note that the decision tree predictions have been used instead of the exact reading of PCs even in cases when the same GPU was used for both model building and autotuning. Therefore, the values on diagonals of tables within Table~\ref{mutitab:hw_port} 
do not copy values from Table~\ref{tab:pc_bias}. For example, the GEMM benchmark, when executed on RTX 2080, gets a speedup of $10.83\times$ with the exact PCs from RTX 2080, whereas the decision tree model from the same GPU limits its speedup to $9.63\times$.

\subsection{Portability across inputs}
\label{eval:input_portability}

\begin{table}[ht]
\centering
\footnotesize
\begin{tabular}{|l|l|l|l|l|}
\hline
	        	& $2048 \times 2048$	& $128 \times 128$	& $16 \times 4096$	& $4096 \times 16$ \\ \hline
$2048 \times 2048$ 	& $8.61\times$ 		& $6.04\times$		& $6.45\times$		& $4.98\times$ \\
$128 \times 128$  	& $2.92\times$  	& $3.17\times$		& $2.71\times$		& $1.72\times$ \\
$16 \times 4096$  	& $3.23\times$ 		& $3.09\times$		& $3.22\times$		& $2.73\times$ \\
$4096 \times 16$  	& $1.5\times$  		& $1.93\times$		& $1.93\times$		& $1.93\times$ \\
\hline
\end{tabular}
\caption{Improvement (in terms of the number of empirical tests) of proposed searcher over random search with the GEMM benchmark. Rows show the sizes used for benchmark execution, columns show the sizes used to create the model of TP and PC relations.}
\label{tab:input_port_gemm}
\end{table}

We further investigated the performance portability for the case of varying input. In our example, we used the GEMM benchmark with significantly varying input matrices:
\begin{itemize}
  \item multiplication of square matrices of size $2048 \times 2048$: this is a typical example of matrix multiplication, which is compute-bound on GPUs;
  \item multiplication of square matrices of size $128 \times 128$: in this case, small matrices do not allow us to utilize the entire GPU easily: the code is rather latency- or strong-scaling bound;
  \item multiplication of highly rectangular matrices (a matrix of size $4096 \times 4096$ multiplied by a matrix of size $16 \times 4096$ and a matrix of size $4096 \times 16$ multiplied by a matrix of size $4096 \times 4096$): in this case, matrix multiplication is memory-bound (because of low arithmetic intensity).
\end{itemize}
We performed the experiment on GTX 1070 with decision tree-based model, which was also built on GTX 1070, but on different inputs. The results of the experiment are shown in Table~\ref{tab:input_port_gemm}. As can be seen, using a different input size typically slightly decreases the speedup. However, the reduction of the number of empirical tests is still significant: for example, autotuning of the compute-bound multiplication of large matrices can be improved more than $6\times$ even if using a model trained on the memory-bound GEMM instance. The improvement with small or highly-rectangular matrices is lower because searching their tuning spaces requires a lower number of searcher steps.

\subsection{Real-time tuning}
\label{sect:realtime}
In this section, we compare the convergence of the proposed profile-based searcher with the convergence of random search in time. Our motivation for this comparison is that random search can converge faster despite requiring more empirical tests because random search has two advantages over the profile-based searcher: it does not collect performance counters (collecting performance counters hinders the kernel execution speed), and it has minimal computational overhead (no model is evaluated). On the other hand, the proposed searcher navigates the search towards faster implementations; therefore, omits evaluation of very slow configurations. Whereas the collection of performance counters is slow especially with slow-running kernels (the kernel runtime dominates over the time of kernel compilation, data copying and selection of new configuration to test), the overhead of the searcher is more significant with vast tuning spaces.

\subsubsection{Experiment setup}

To test the convergence time of the searchers, we have used a machine equipped with NVIDIA GeForce RTX 2080, Intel Core i7-8700, 32\,GB RAM, Ubuntu 18.04.3, NVIDIA driver 418.39 and CUDA 10.1. Each benchmark has been executed a hundred times. The model has been created using data obtained with GeForce GTX 1070. Therefore, our test simulates a situation when the user acquires a brand new GPU and has autotuning data from an older GPU. 

We have performed autotuning with persistent data on GPU and no comparison to the reference computation. Such a setting is typical for dynamic autotuning. It improves the results of the random searcher, as there is no overhead of data movement and comparison, which could hide the slower execution of kernels while gathering performance counters. On the other hand, we have selected the size of the kernels' inputs such that the GPU is fully occupied, but the kernels are not running longer than necessary (we consider 1-10 milliseconds as a suitable kernel runtime for our experiments). This choice, on the other hand, improves the results of the proposed searcher, because profiled kernels do not run for too long. In two experiments, we switch on the comparison to the reference kernel and use larger kernel input to demonstrate the described effects.

In this evaluation, we show the searcher's convergence as a graph of the average kernel runtime at each second of autotuning. When a long-running kernel is selected in the first iteration of the search, the benchmark has no finished kernel for several seconds after its execution (this is especially the case when gathering performance counters). We decided to draw the graph from the time when all 100 experiments have at least one finished kernel and therefore its time is known. Such a solution prevents the results of the proposed searcher from being artificially improved, \eg{}, by only plotting the runtime of the finished (and thus fast) kernels during the first seconds of measurements.

\subsubsection{Results}

\begin{figure}[t]
  \centering
  \includegraphics[width=.85\hsize]{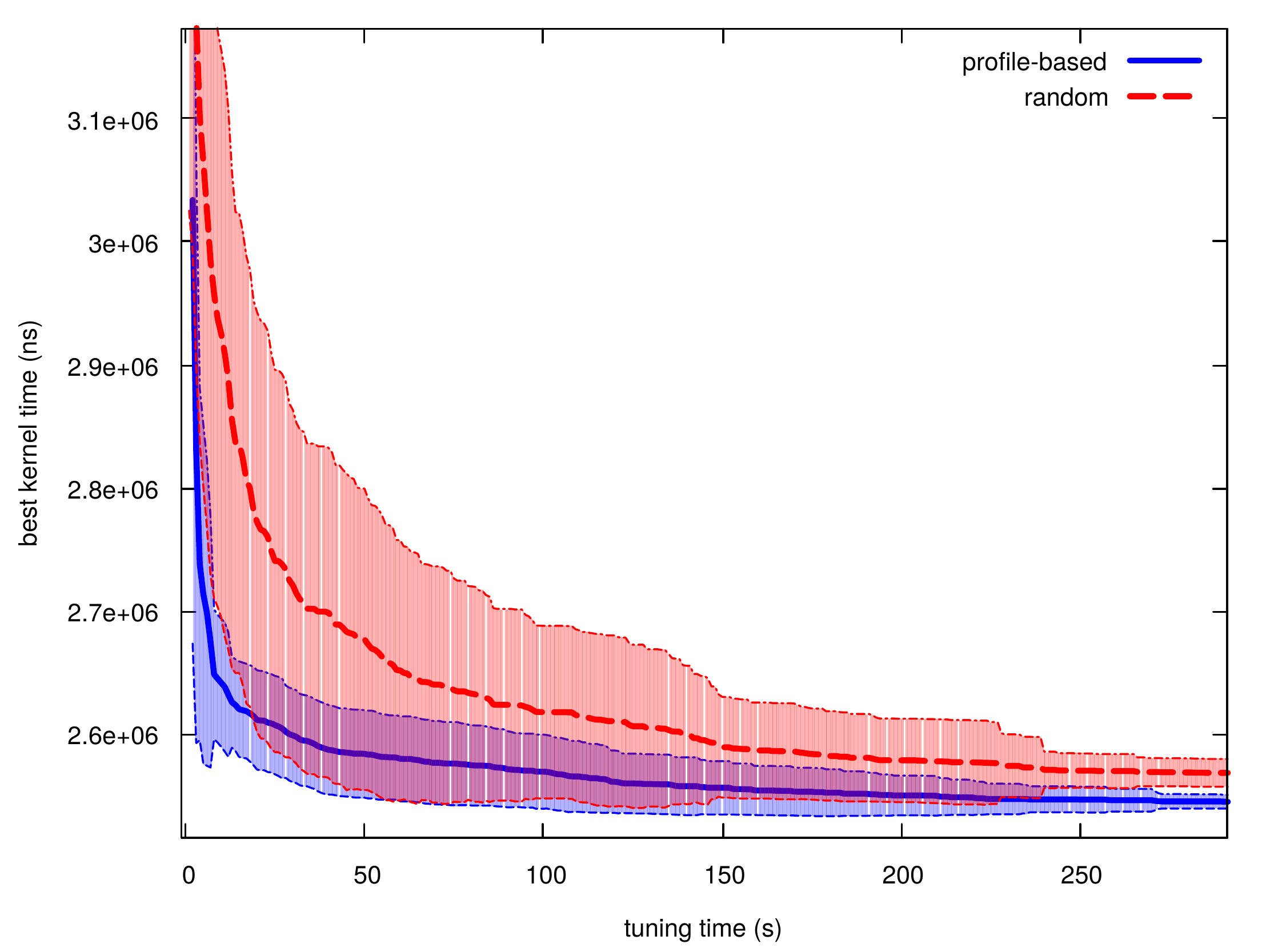}
  \caption{Convergence of GEMM, $2048 \times 2048 \times 2048$, GTX 2080, model from GTX 1070. The solid line shows the average, the transparent area shows the standard deviation.}
  \label{fig:gemm_convergence}
\end{figure}

\begin{figure}[t]
  \centering
  \includegraphics[width=.85\hsize]{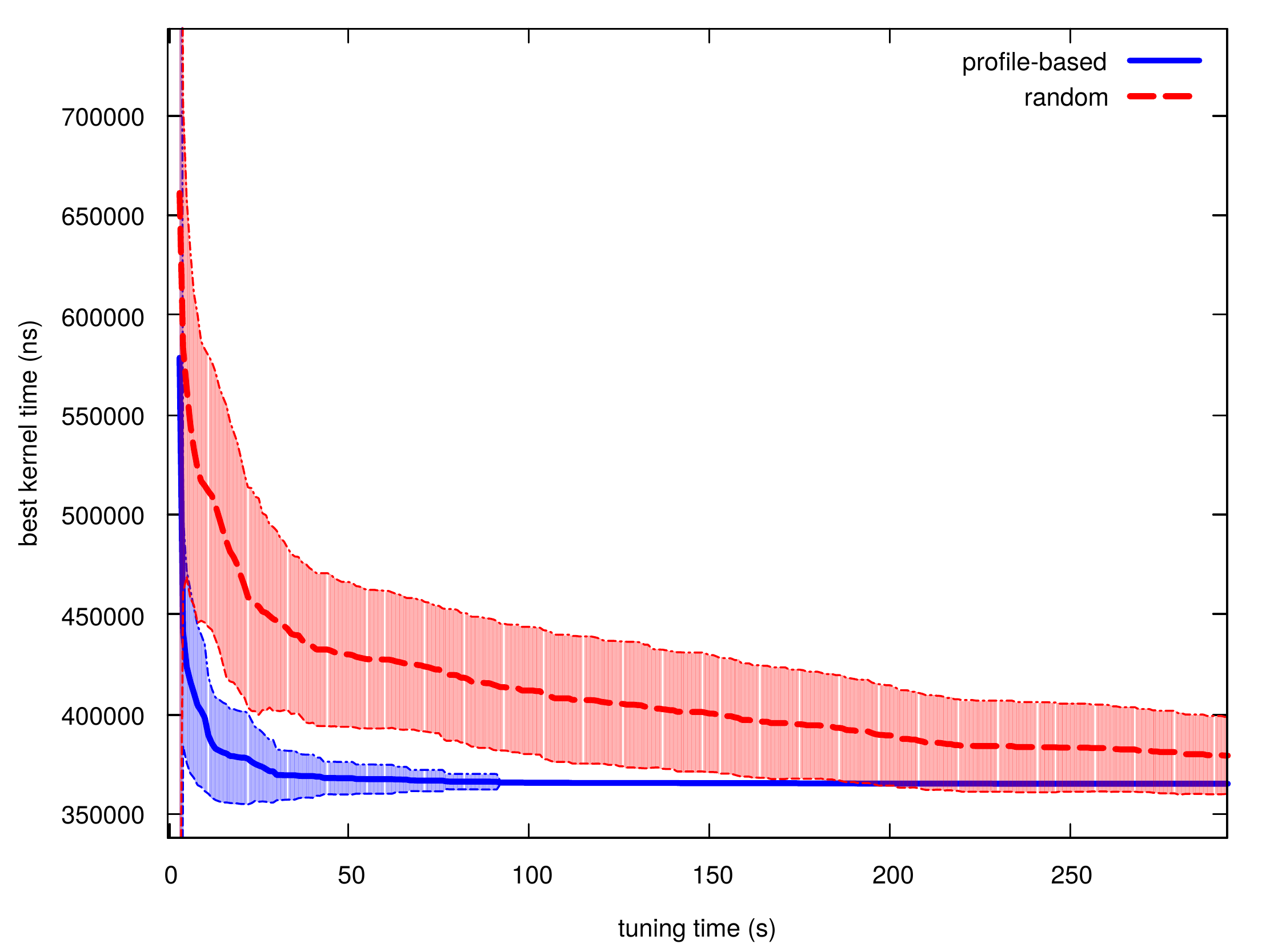}
  \caption{Convergence of Convolution, $4096 \times 4096$, GTX 2080, model from GTX 1070. The solid line shows the average, the transparent area shows the standard deviation.}
  \label{fig:conv_convergence}
\end{figure}

The GEMM and Convolution benchmarks are the most difficult cases to autotune (see Table~\ref{tab:random_iters}). Therefore, improving the speed of convergence is more important for those benchmarks. The GEMM benchmark was run using square matrices of size $2048 \times 2048$, the Convolution benchmark using a 2D array of size $4096 \times 4096$. The comparison of their convergence can be seen in Figure~\ref{fig:gemm_convergence} and Figure~\ref{fig:conv_convergence}, respectively. In both cases, the proposed searcher brings significant convergence speedup.

\begin{figure}[t]
  \centering
  \includegraphics[width=.49\hsize]{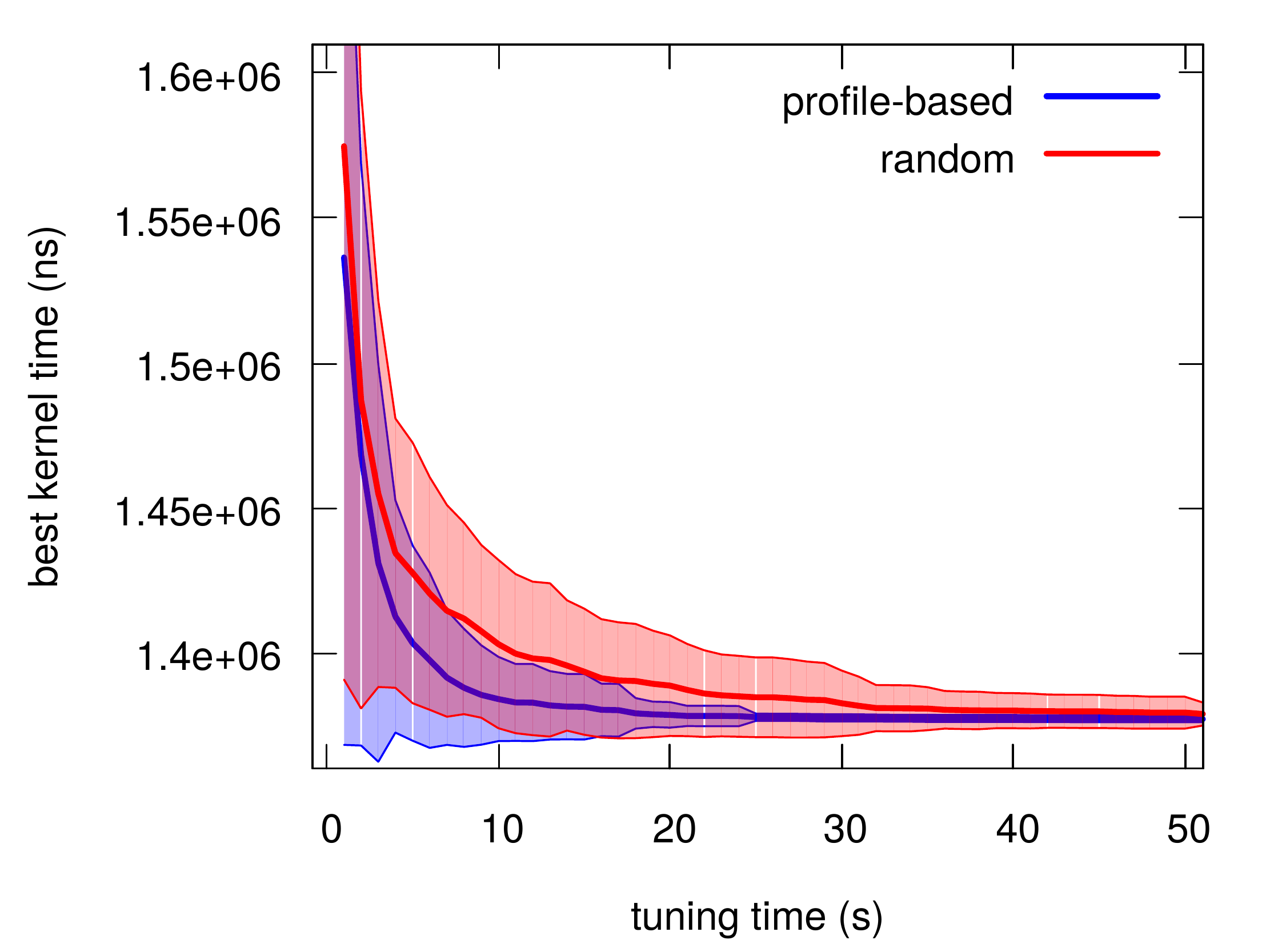}
  \includegraphics[width=.49\hsize]{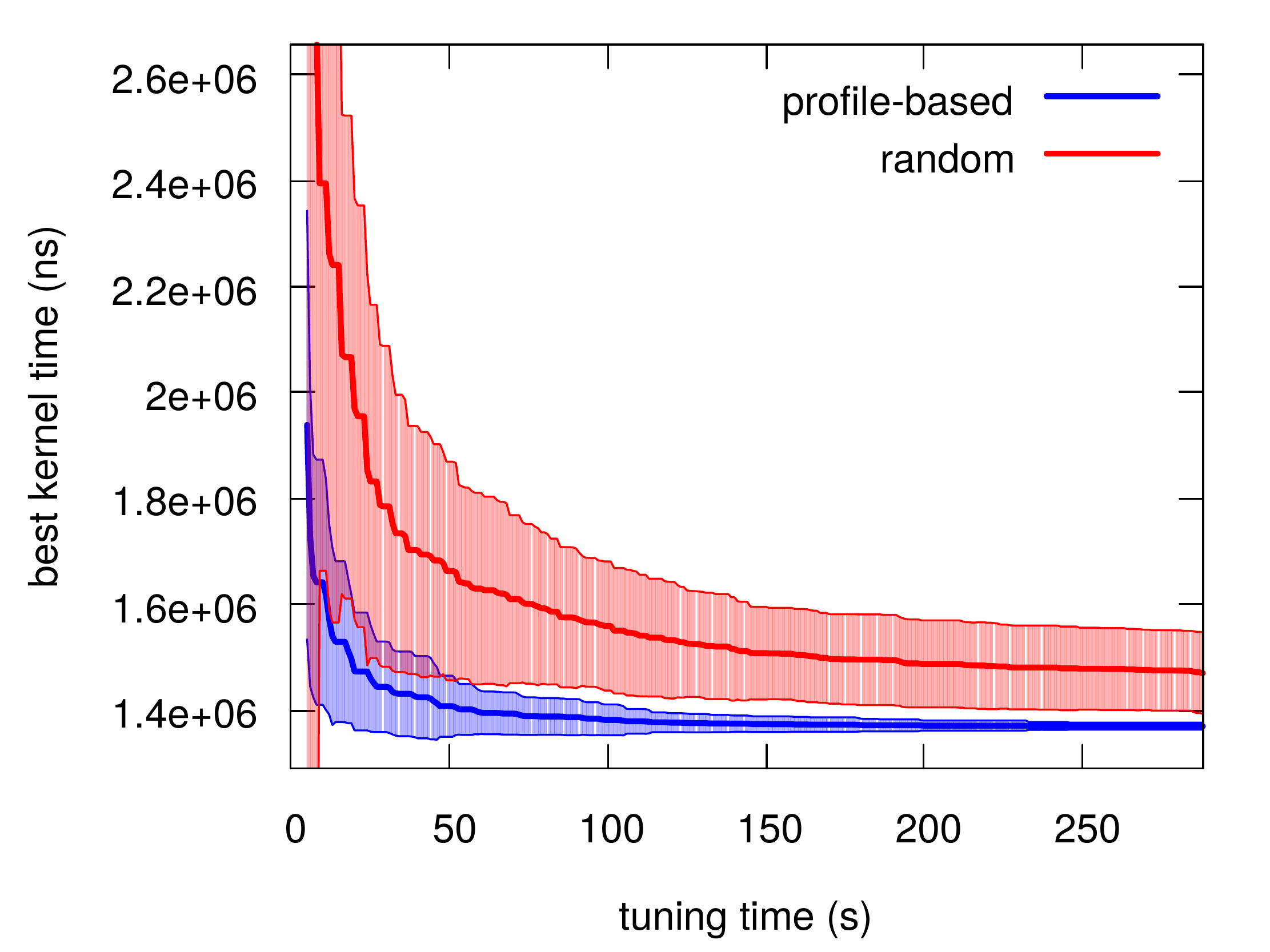}
  \caption{Convergence of Matrix transposition, $8192 \times 8192$, GTX 2080, model from GTX 1070. Left: tuning configured to not check kernel results. Right: tuning configured to check kernel results. The solid line shows the average, the transparent area shows the standard deviation.}
  \label{fig:mtran_convergence}
\end{figure}

The Matrix transposition benchmark was tested with matrix size $8192 \times 8192$. Although the proposed searcher requires fewer empirical tests, the convergence in time is not significantly faster comparing to the random searcher -- see Figure~\ref{fig:mtran_convergence}. We also tested another scenario: the tuning was set to check results against the reference implementation, which is common during offline tuning. In such a case, the matrices are copied into host memory and checked against the reference, which adds a constant overhead for each empirical test. In this case, the overhead of gathering performance counters is less visible, and the proposed searcher converges significantly faster (Figure~\ref{fig:mtran_convergence}).

\begin{figure}[t]
  \centering
  \includegraphics[width=.49\hsize]{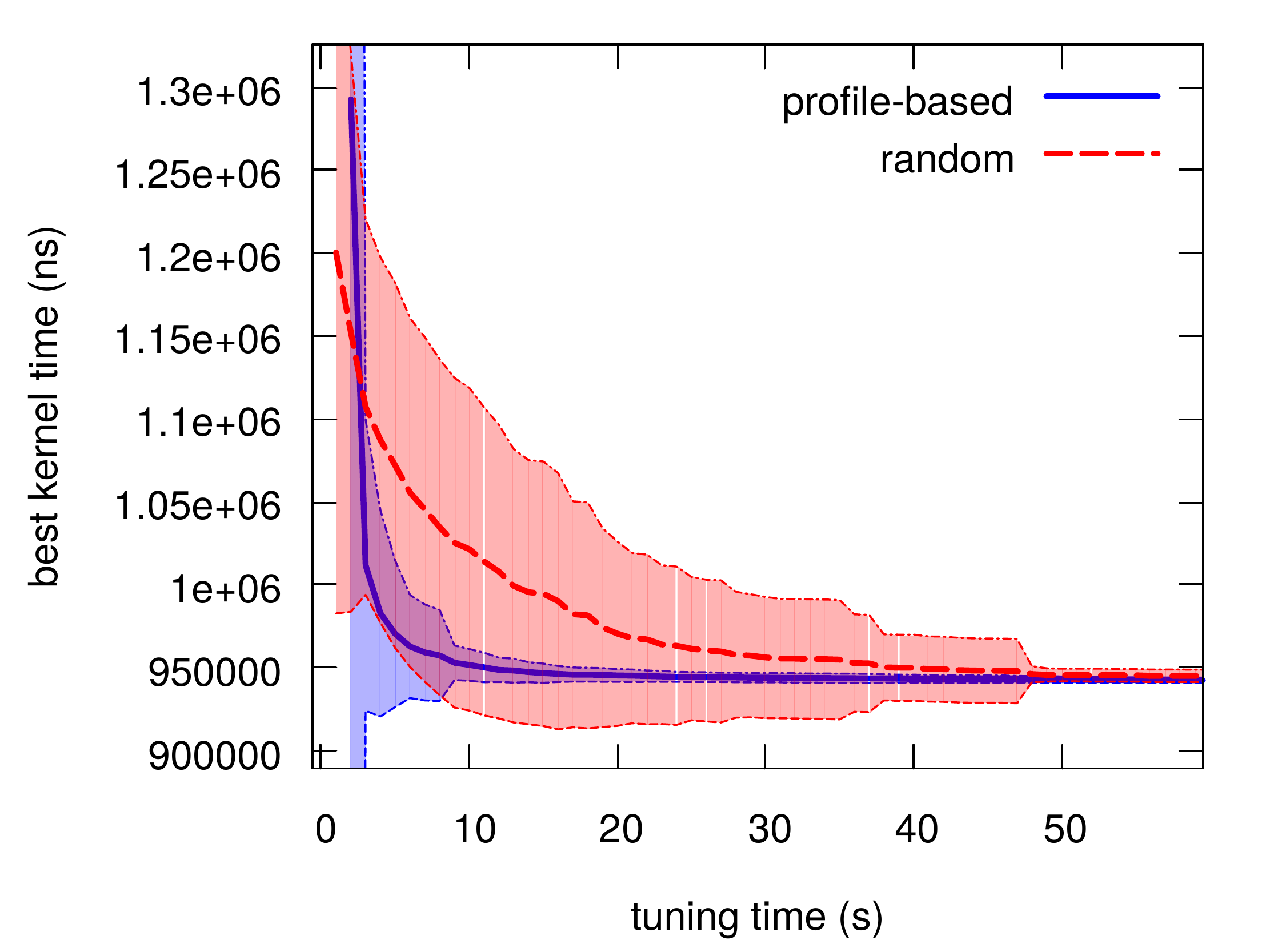}
  \includegraphics[width=.49\hsize]{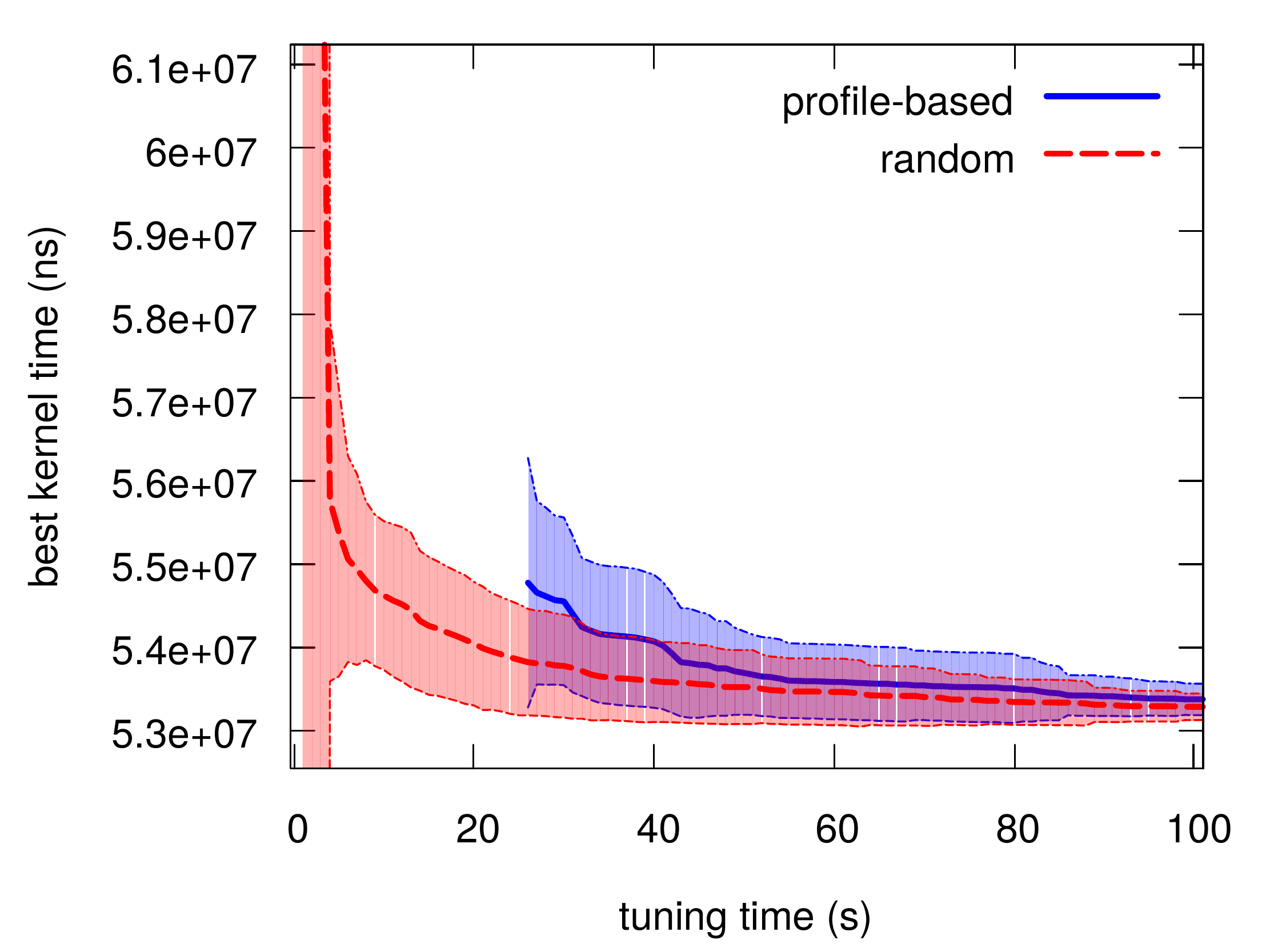}
  \caption{Convergence of n-body on GTX 2080, model from GTX 1070. Left: tuning with 16,384 bodies. Right: tuning with 131,072 bodies. The solid line shows the average, the transparent area shows the standard deviation.}
  \label{fig:nbody_convergence}
\end{figure}

The n-body benchmark was tested with 16,384 bodies. Although the overall number of searcher iterations is not high even for the random search, the proposed searcher converges significantly faster -- see Figure~\ref{fig:nbody_convergence}. We also tested a situation with tuning executed on a much bigger problem instance of 131,072 bodies (note that n-body is in $\mathcal{O}(n^2)$, where $n$ is the number of bodies). In such a setup, kernels run much longer and profiling overhead included in the proposed searcher makes it slower than random searcher (see Figure~\ref{fig:nbody_convergence}).

\begin{figure}[t]
  \centering
  \includegraphics[width=.85\hsize]{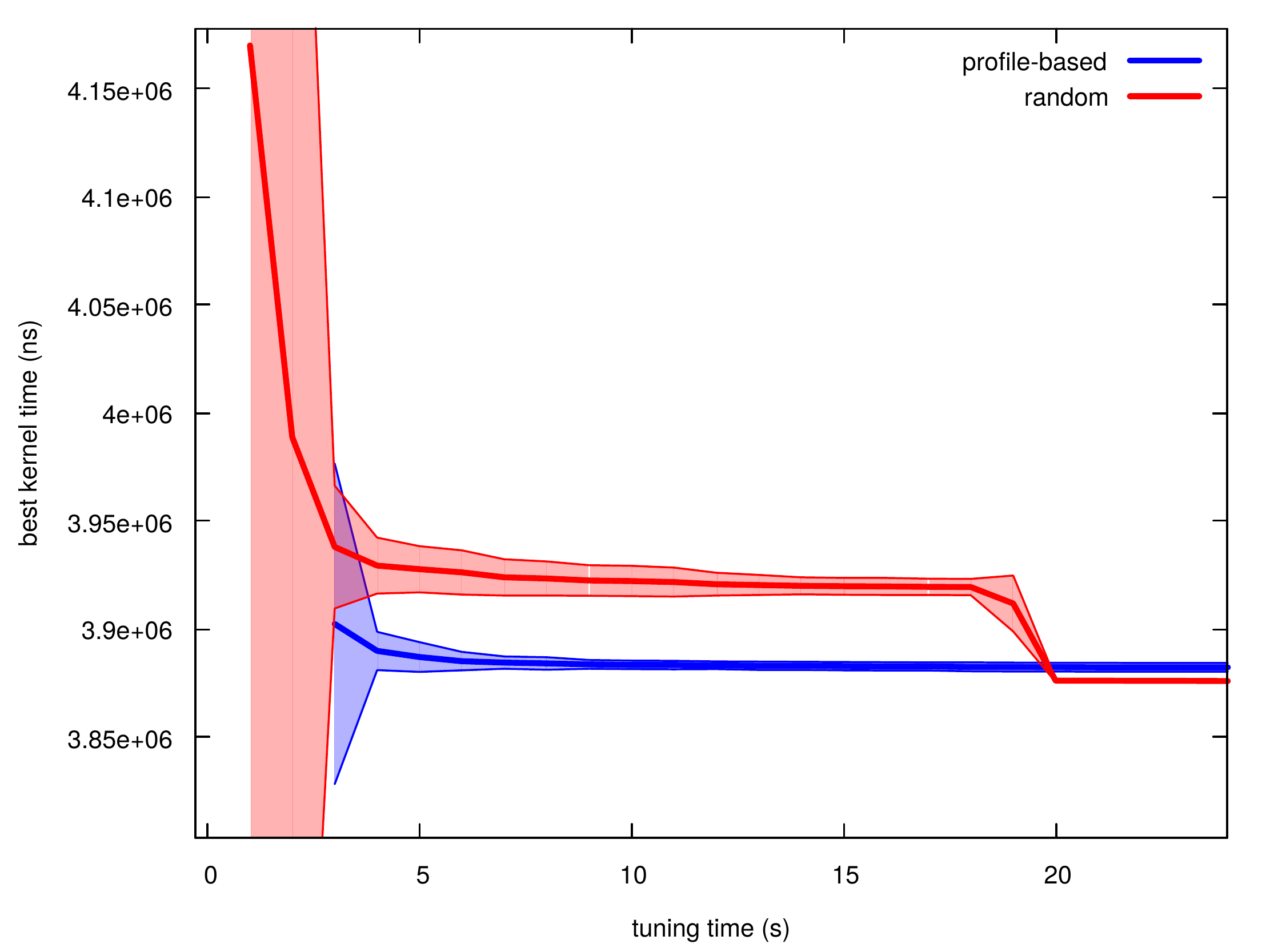}
  \caption{Convergence of Coulomb Sum, grid $256 \times 256 \times 256$, 256 atoms, GTX 2080, model from GTX 1070. The solid line shows the average, the transparent area shows the standard deviation.}
  \label{fig:coulomb_convergence}
\end{figure}

The Coulomb sum benchmark uses a grid of $256 \times 256 \times 256$ cells and $256$ atoms. It converges very quickly, even with the random searcher. Figure~\ref{fig:coulomb_convergence} shows that the proposed searcher needs some time for initial profiling. Then, it converges very quickly, whereas random searcher matches its performance after 20 seconds of tuning. However, both searchers converge close to the optimum in less than 5 seconds. Therefore, the choice of the searcher is not important here.

\begin{figure}[t]
  \centering
  \includegraphics[width=.85\hsize]{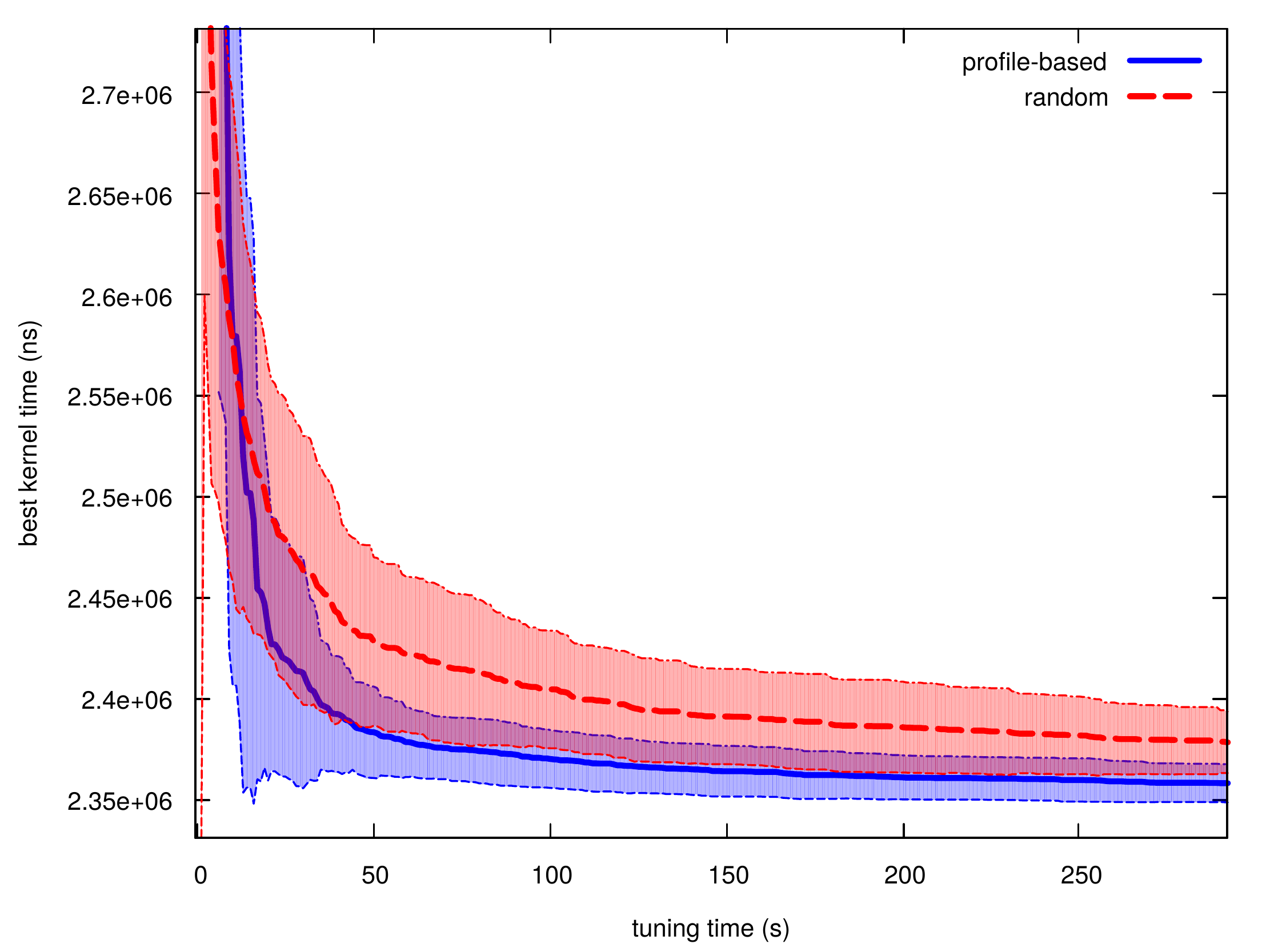}
  \caption{Convergence of GEMM full, $2048 \times 2048 \times 2048$, GTX 2080, model from GEMM on GTX 1070. The solid line shows the average, the transparent area shows the standard deviation.}
  \label{fig:gemm_full_convergence}
\end{figure}

The setup of this experiment also allows us to benchmark the GEMM full benchmark. As we have not performed exhaustive search of GEMM full, the model was created from the tuning space of the GEMM benchmark, measured on GeForce GTX 1070. Note that the GEMM benchmark does not cover the tuning space of the GEMM full benchmark completely -- it lacks four tuning parameters and uses less then 3\% of GEMM full configurations. However, Figure~\ref{fig:gemm_full_convergence} shows that the proposed searcher converges much faster than random searcher with the GEMM full benchmark: the proposed searcher requires 61 seconds to match the same results as the random searcher reaches after 300 seconds. This result can be further improved by optimizing the implementation of the proposed searcher: with the GEMM full tuning space, it requires significant time to score tuning configurations, resulting in $3\times$ longer time per empirical test.

\subsection{Comparison to Basin Hopping}

So far, we have compared our proposed profile-based searcher to random search. However, the Basin Hopping optimization has recently been shown to perform better than many other optimization methods~\cite{vanwerkhoven2018kernel}. Therefore, we also compare our searcher (with the settings described in Section~\ref{sect:realtime}) to the Basin Hopping searcher implemented in Kernel Tuner~\cite{vanwerkhoven2018kernel}. 

\subsubsection{Experiment setup}
We used the same machine as for KTT (see Section~\ref{sect:realtime}) to test Kernel Tuner. We have re-implemented KTT examples for Kernel Tuner and autotuned them with Basin Hopping optimization in default settings~\footnote{We are aware that the parameters of the Basin Hopping method can be tuned towards improving search convergence speed. However, the optimal values of those parameters are not known \textit{a priori}, so they cannot be tuned before autotuning for new hardware or input is finished. Therefore, we kept the default settings for a fair comparison.}.

During the testing of Kernel Tuner, we have observed that it works slower than KTT, even when the random searcher is used. We suppose that the source of its slower speed is twofold. First, Kernel Tuner is implemented in Python, whereas KTT is a C++ native application. Second, Kernel Tuner executes each kernel three times to get a better timing precision. Although KTT only executes each kernel once, it gives more consistent results: when executed multiple times, the KTT timing is more stable than performance times reported by Kernel Tuner. We suppose this is caused by the overhead of the python-based Kernel Tuner implementation. We also found that the measurement of Kernel Tuner systematically reports slightly higher kernel runtimes compared to KTT. Therefore, we have normalized performance times to be comparable: we have done an exhaustive search of the tuning space and multiplied the times reported by Kernel Tuner by the ratio of the best time measured by KTT and the best time measured by Kernel Tuner. Therefore, both tuning systems converge at the same kernel times in our figures.

As the comparison of timing is affected by the slower execution of Kernel Tuner, we have also compared the number of searcher steps here. The number of steps allows us to deduce how well Basin Hopping navigates the tuning space compared to random search and the proposed search based on performance counters, without the handicap of slower Kernel Tuner execution.

\subsubsection{Results}

\begin{figure}[t]
  \centering
  \includegraphics[width=.49\hsize]{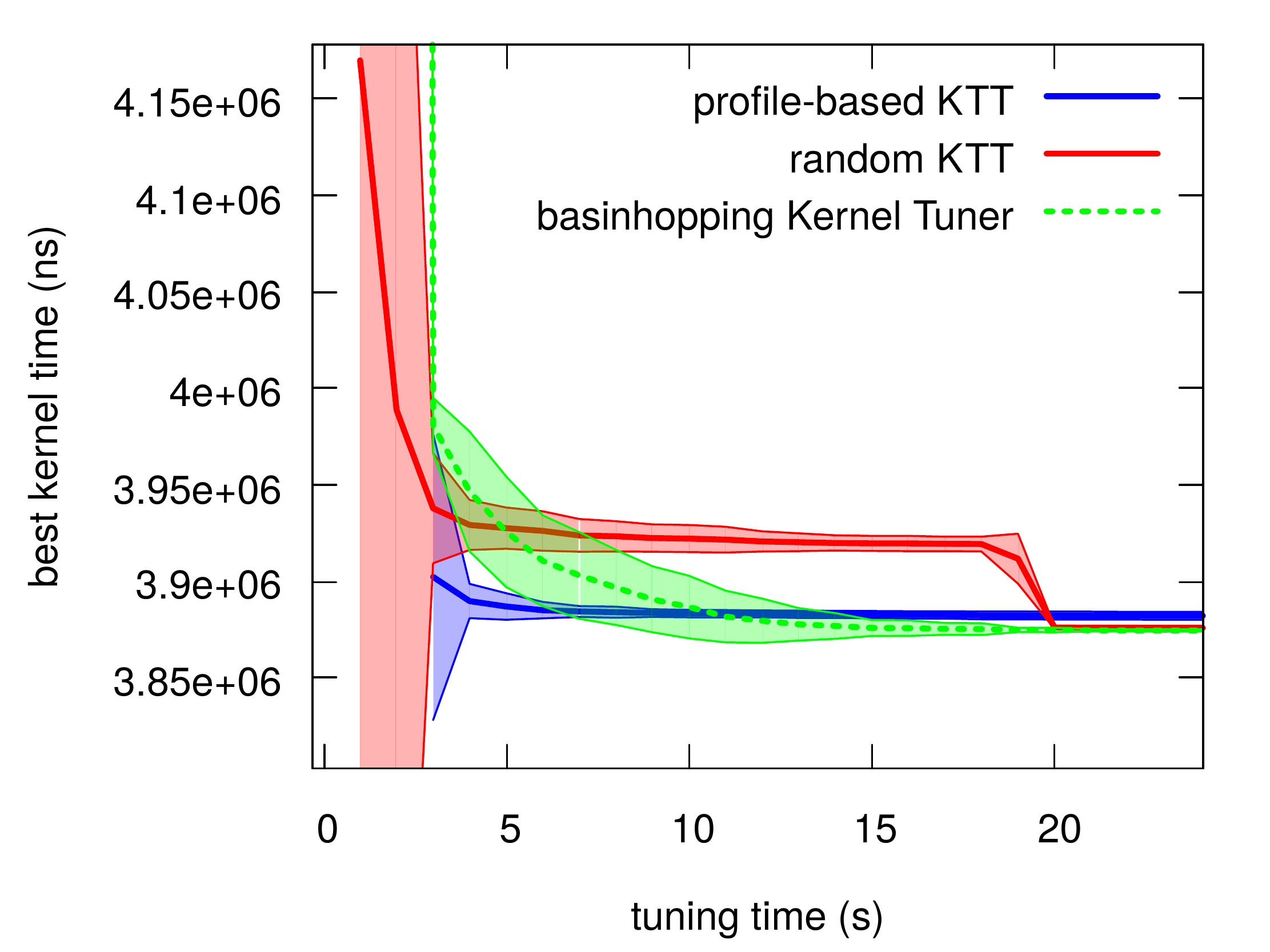}
  \includegraphics[width=.49\hsize]{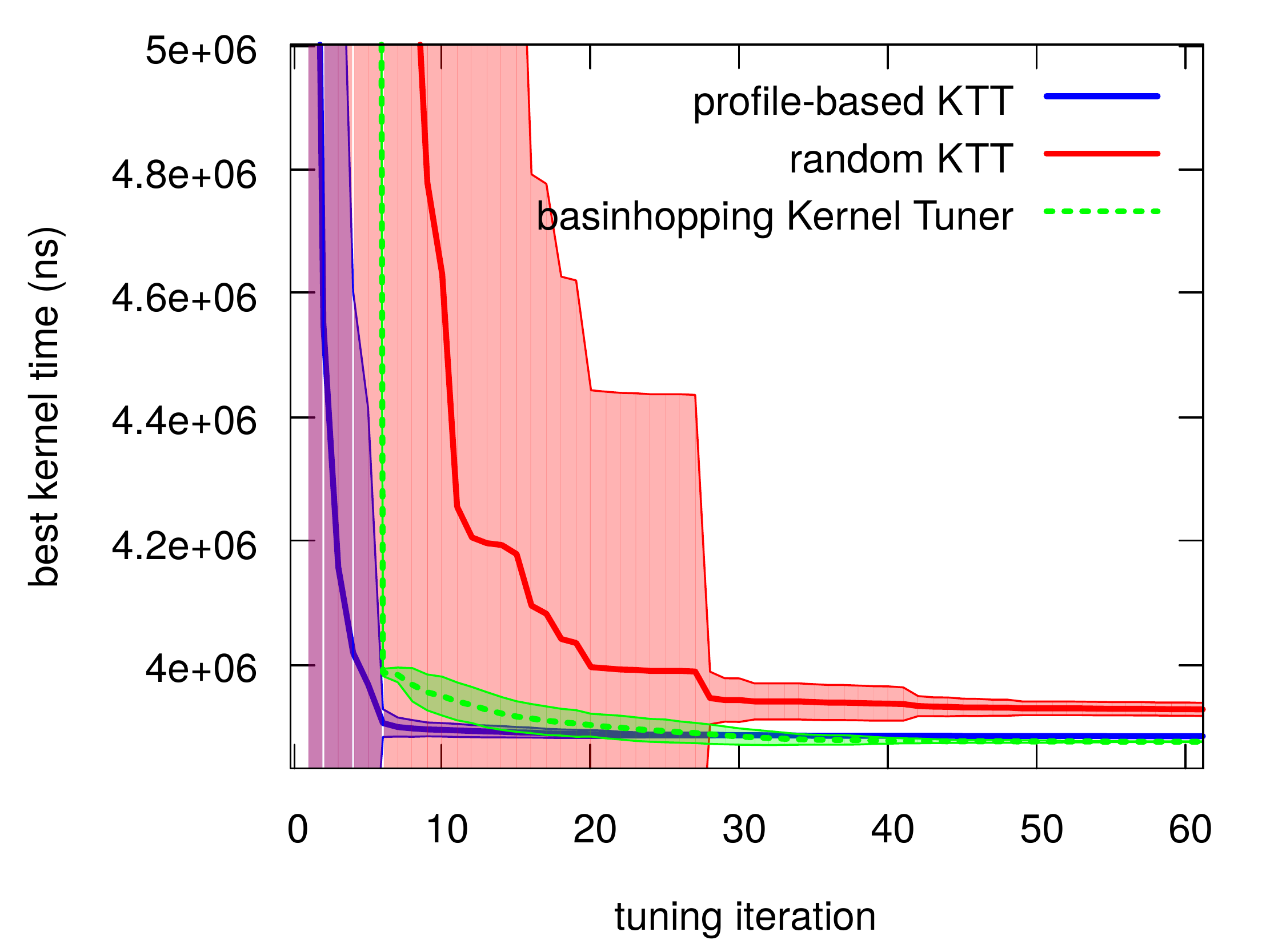}
  \caption{Convergence of the Coulomb sum benchmark using KTT and Kernel Tuner. Left: convergence speed in time. Right: comparison of iterations (empirical tests). The solid line shows the average, transparent area shows the standard deviation.}
  \label{fig:gemm_coulomb}
\end{figure}

Figure~\ref{fig:gemm_coulomb} shows a comparison of the proposed searcher implemented in KTT and Basin Hopping implemented in Kernel Tuner with Coulomb sum benchmark. Comparing convergence times, the proposed searcher converges faster during the first 10 seconds. After 10 seconds, it is slightly outperformed by Basin Hopping. When we compare the number of empirical tests, the proposed searcher behaves similarly to Basin Hopping.

\begin{figure}[t]
  \centering
  \includegraphics[width=.49\hsize]{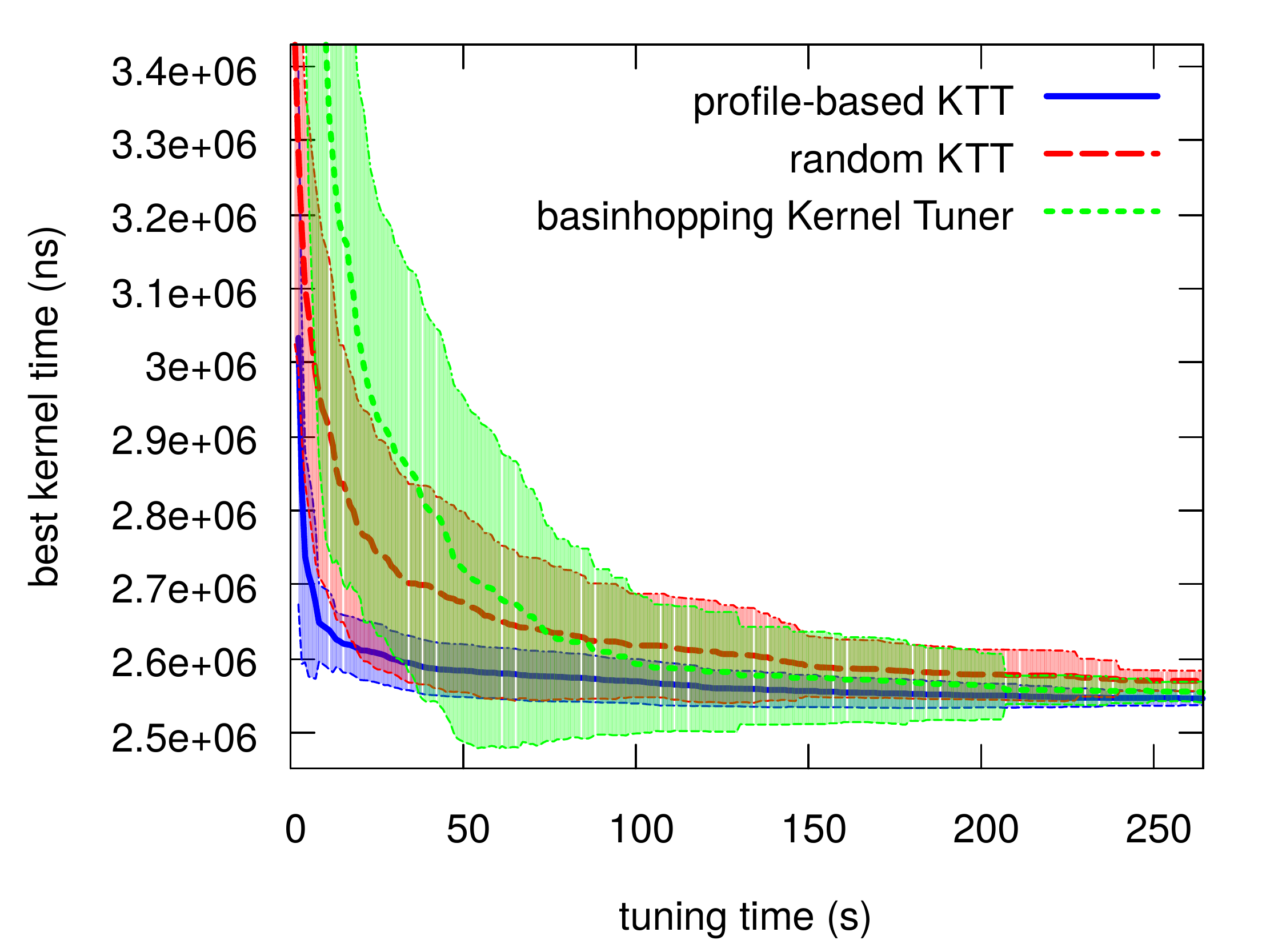}
  \includegraphics[width=.49\hsize]{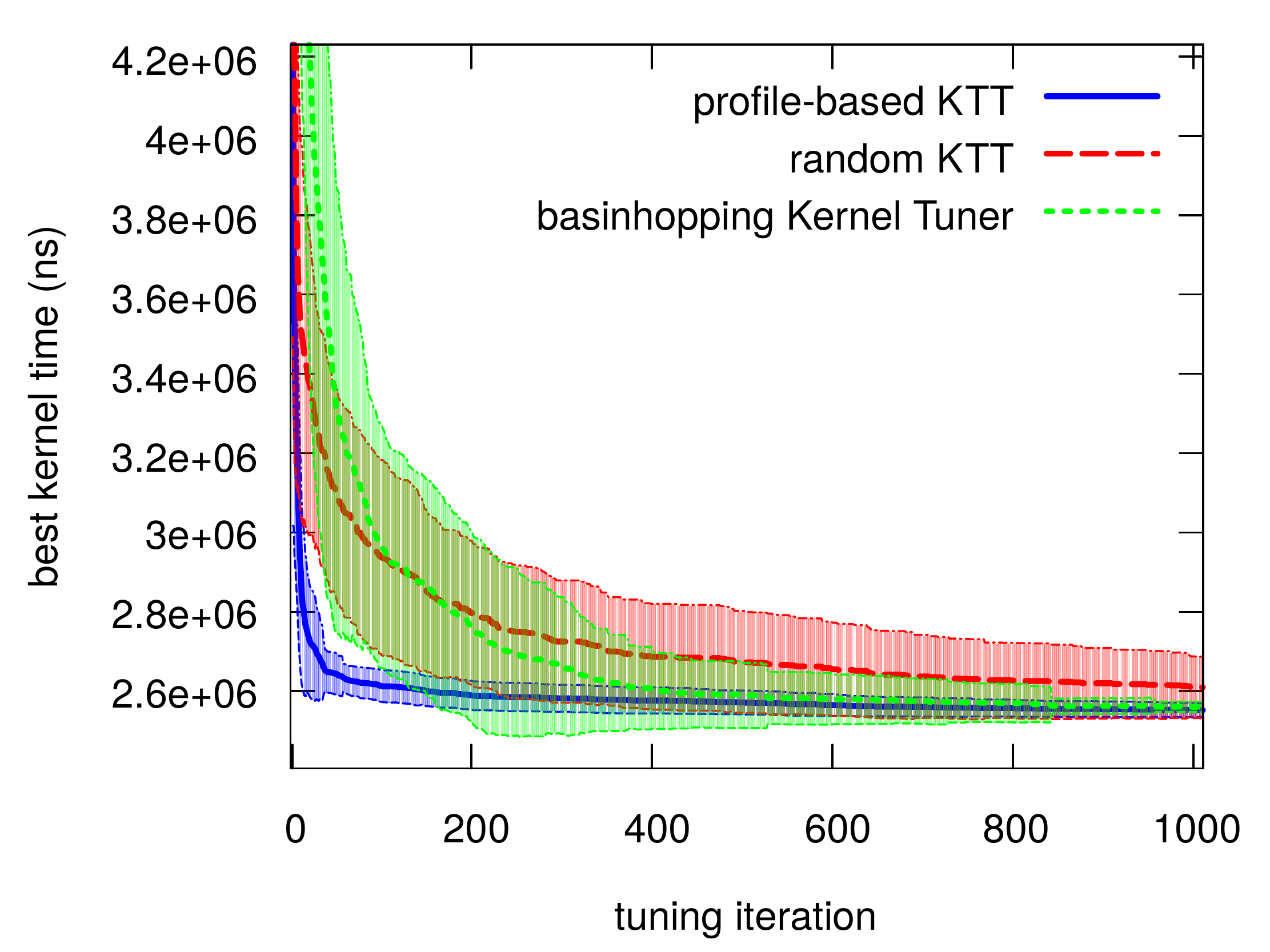}
  \caption{Convergence of the GEMM benchmark using KTT and Kernel Tuner. Left: convergence speed in time. Right: comparison of iterations (empirical tests). The solid line shows the average, the transparent area shows the standard deviation.}
  \label{fig:gemm_kt}
\end{figure}

The GEMM benchmark is compared in Figure~\ref{fig:gemm_kt}. It can be seen that the Basin Hopping optimization implemented in Kernel Tuner converges more slowly than random searcher during the first 70 seconds and then it outperforms random search slightly. However, the reason for such a result of Basin Hopping is related to the slower execution in Kernel Tuner. When we compare the number of empirical tests, Basin Hopping converges to the optimum much faster than random searcher after the first 200 iterations. The proposed profile-based searcher uses significantly fewer empirical tests than both Random and Basin Hopping and outperforms them in both the number of empirical tests and the convergence time.

\begin{figure}[t]
  \centering
  \includegraphics[width=.49\hsize]{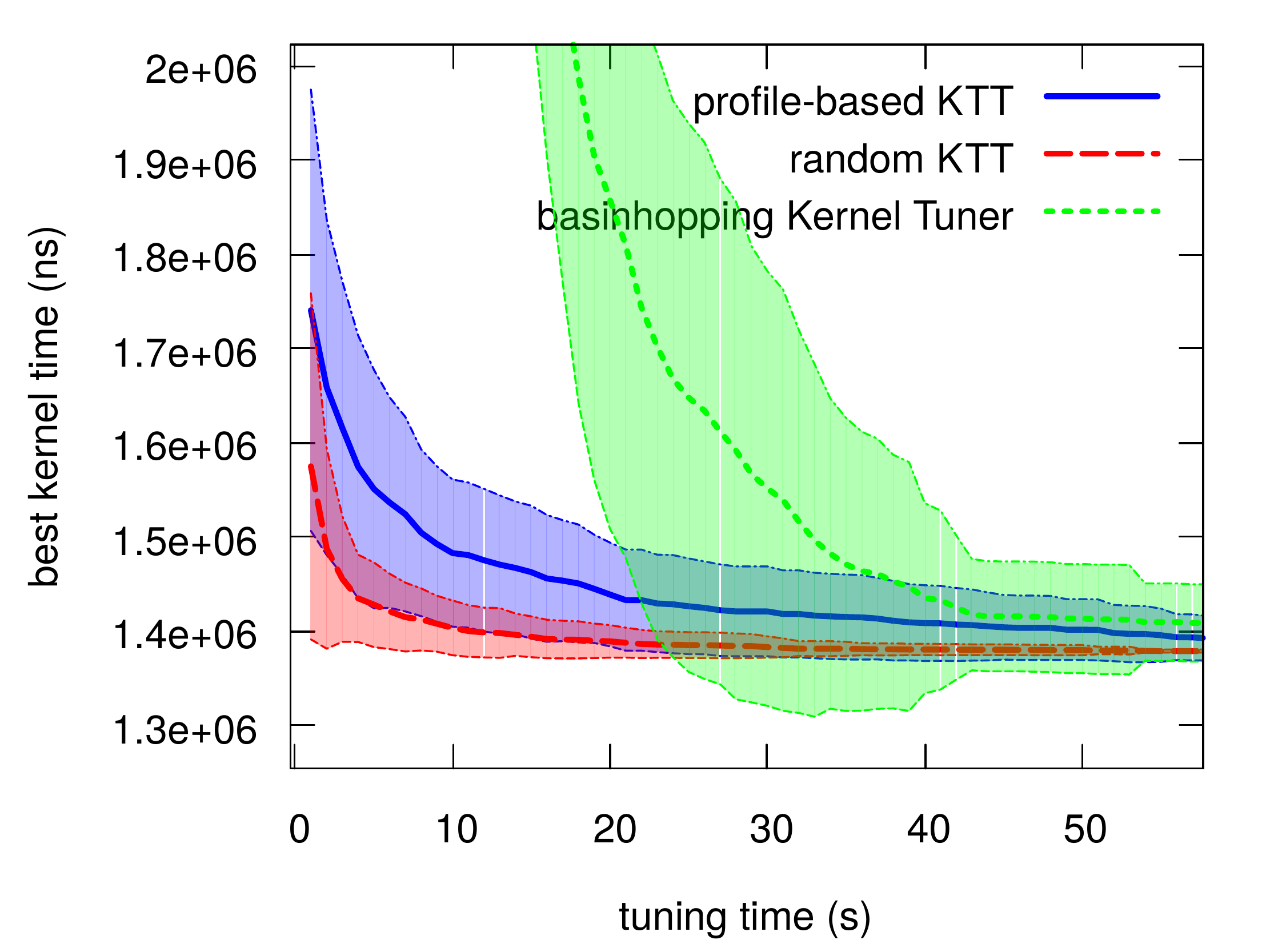}
  \includegraphics[width=.49\hsize]{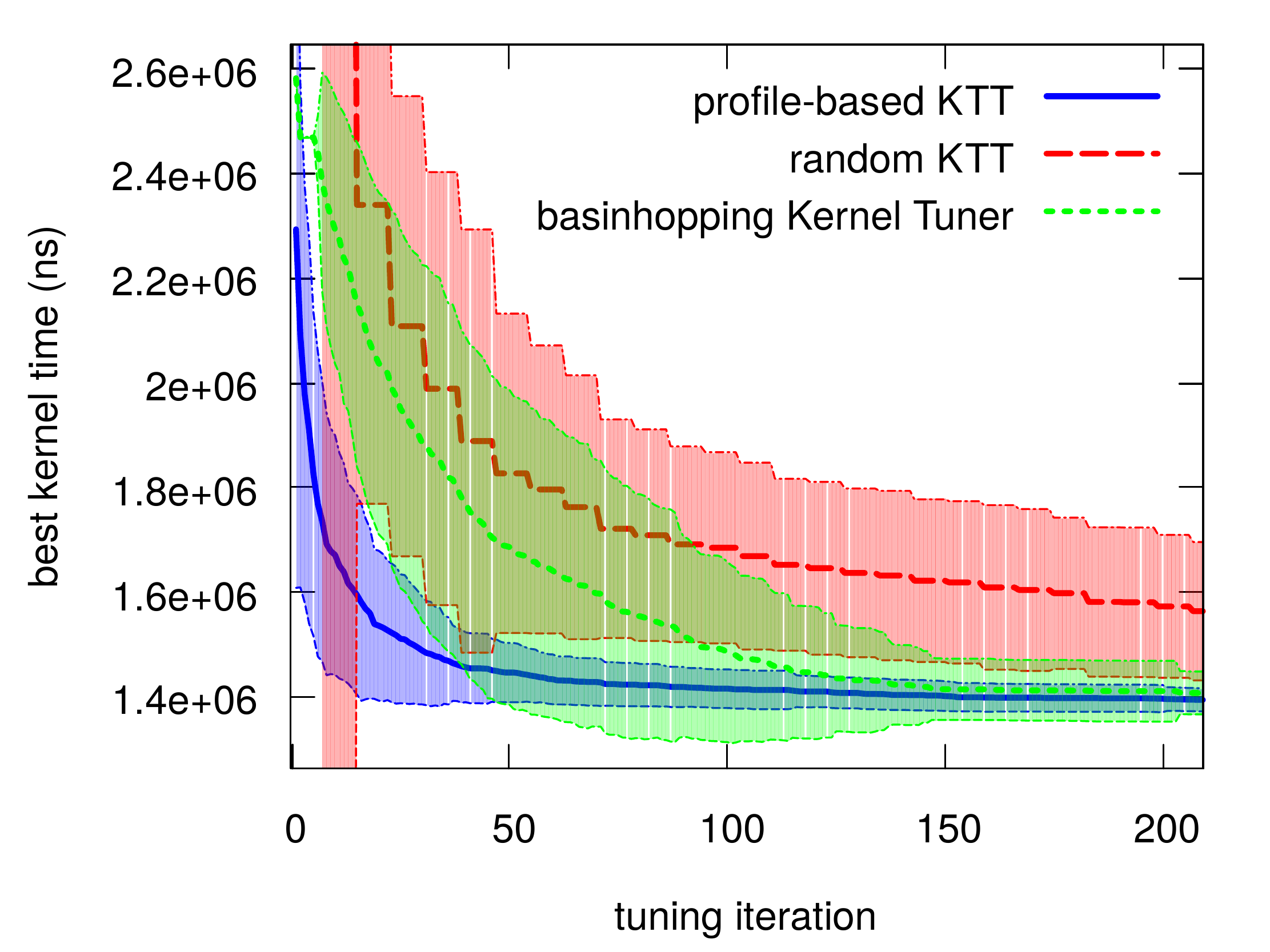}
  \caption{Convergence of the Matrix transposition benchmark using KTT and Kernel Tuner. Left: convergence speed in time. Right: comparison of iterations (empirical tests). The solid line shows the average, transparent area shows the standard deviation.}
  \label{fig:mtran_kt}
\end{figure}

With the Matrix transposition benchmark, the slowdown of Kernel Tuner is more visible, see Figure~\ref{fig:mtran_kt}. Kernel Tuner requires about 16 seconds to start tuning, probably because of complicated constraints pruning the tuning space. After initialization, it converges quickly, but cannot outperform random or proposed searcher. When we compare the number of empirical tests, Basin Hopping again converges near the optimum much faster than the random searcher. However, the proposed profile-based searcher uses significantly fewer empirical tests.

\begin{figure}[t]
  \centering
  \includegraphics[width=.49\hsize]{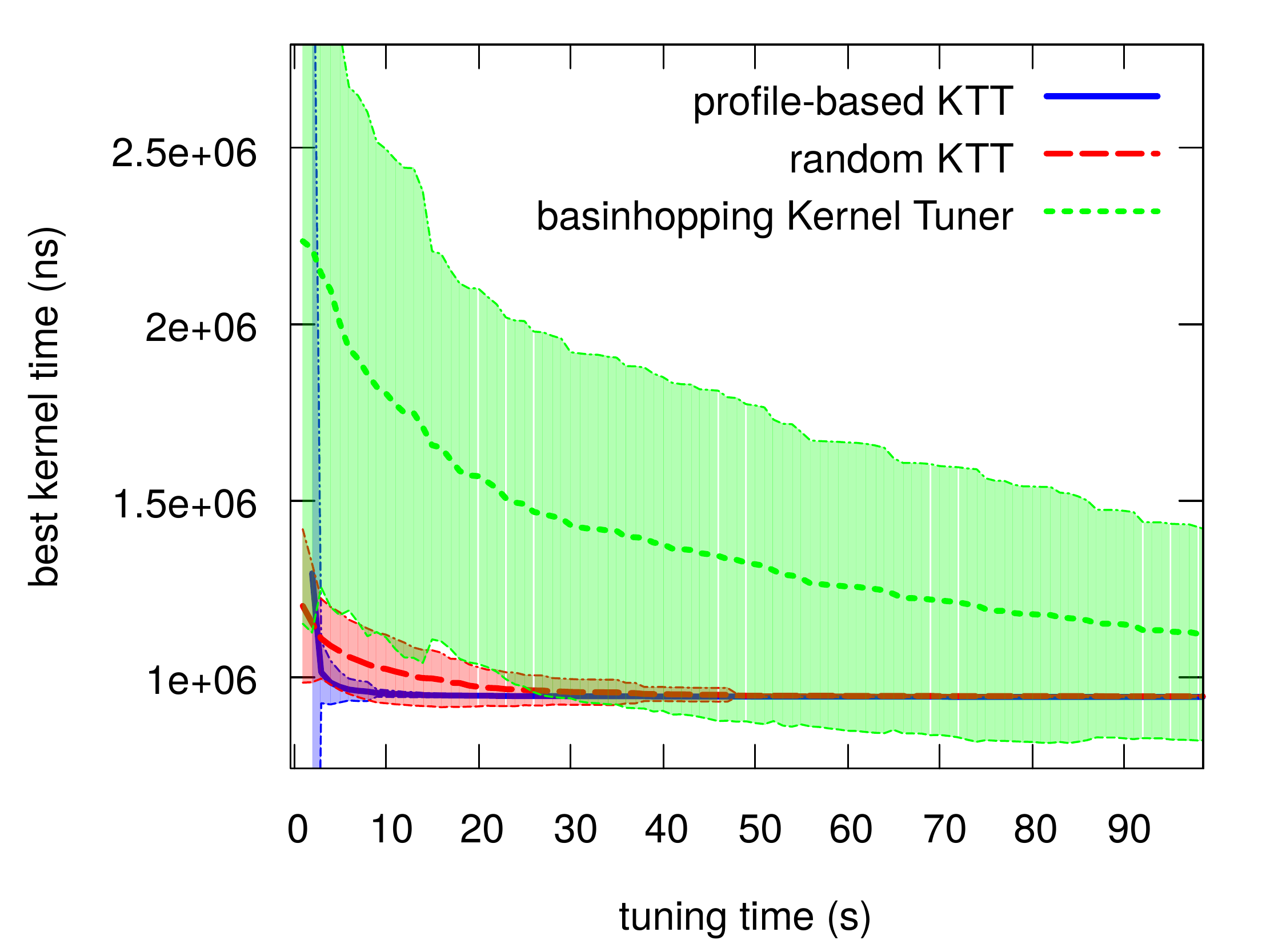}
  \includegraphics[width=.49\hsize]{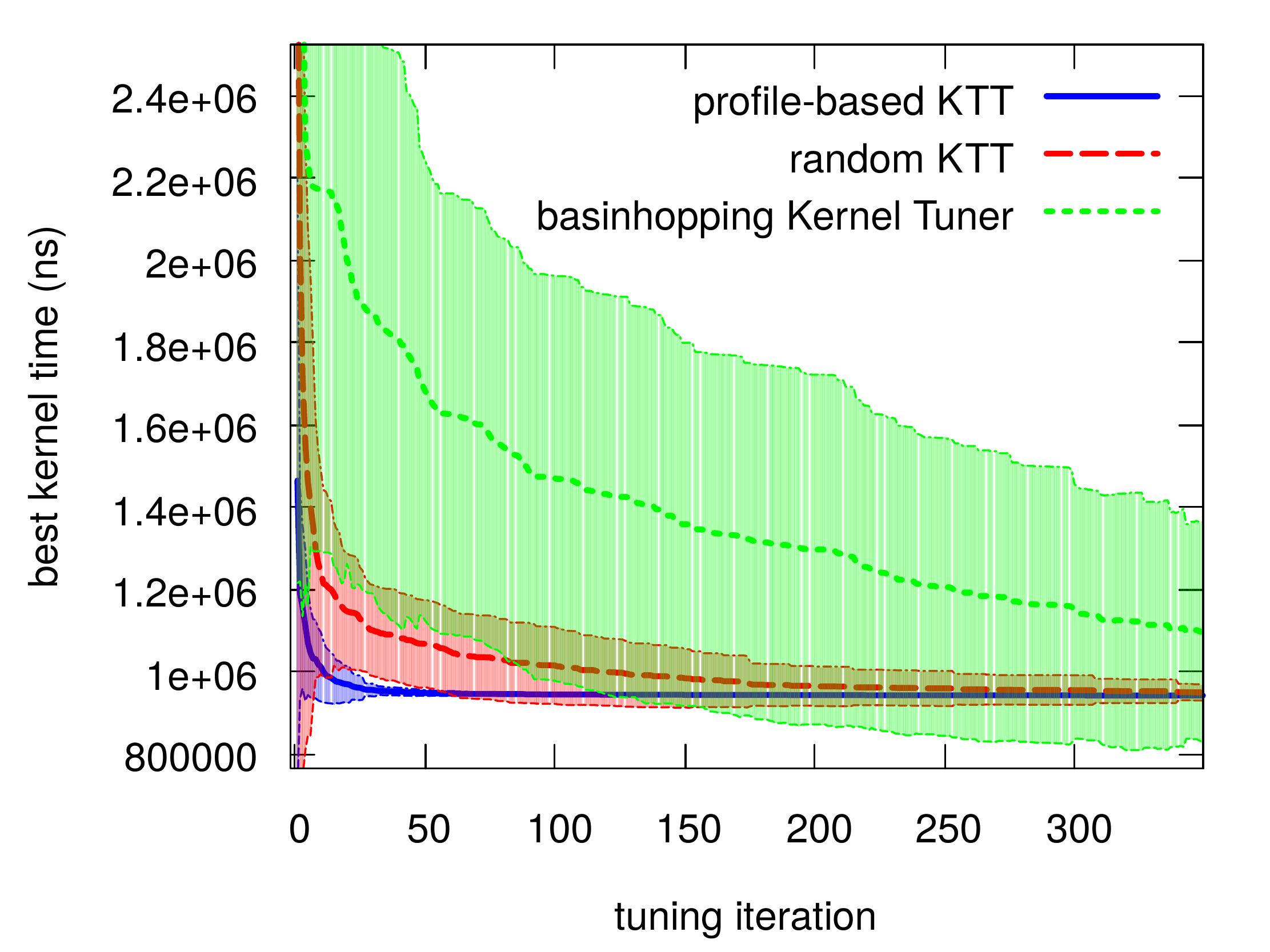}
  \caption{Convergence of the n-body benchmark using KTT and Kernel Tuner. Left: convergence speed in time. Right: comparison of iterations (empirical tests). The solid line shows the average, transparent area shows the standard deviation.}
  \label{fig:nbody_kt}
\end{figure}

With the n-body benchmark, the Basin Hopping method does not converge well, see Figure~\ref{fig:nbody_kt}. Comparing both convergence time and number of empirical tests, it performs worse than random searcher and the proposed searcher.

\begin{figure}[t]
  \centering
  \includegraphics[width=.49\hsize]{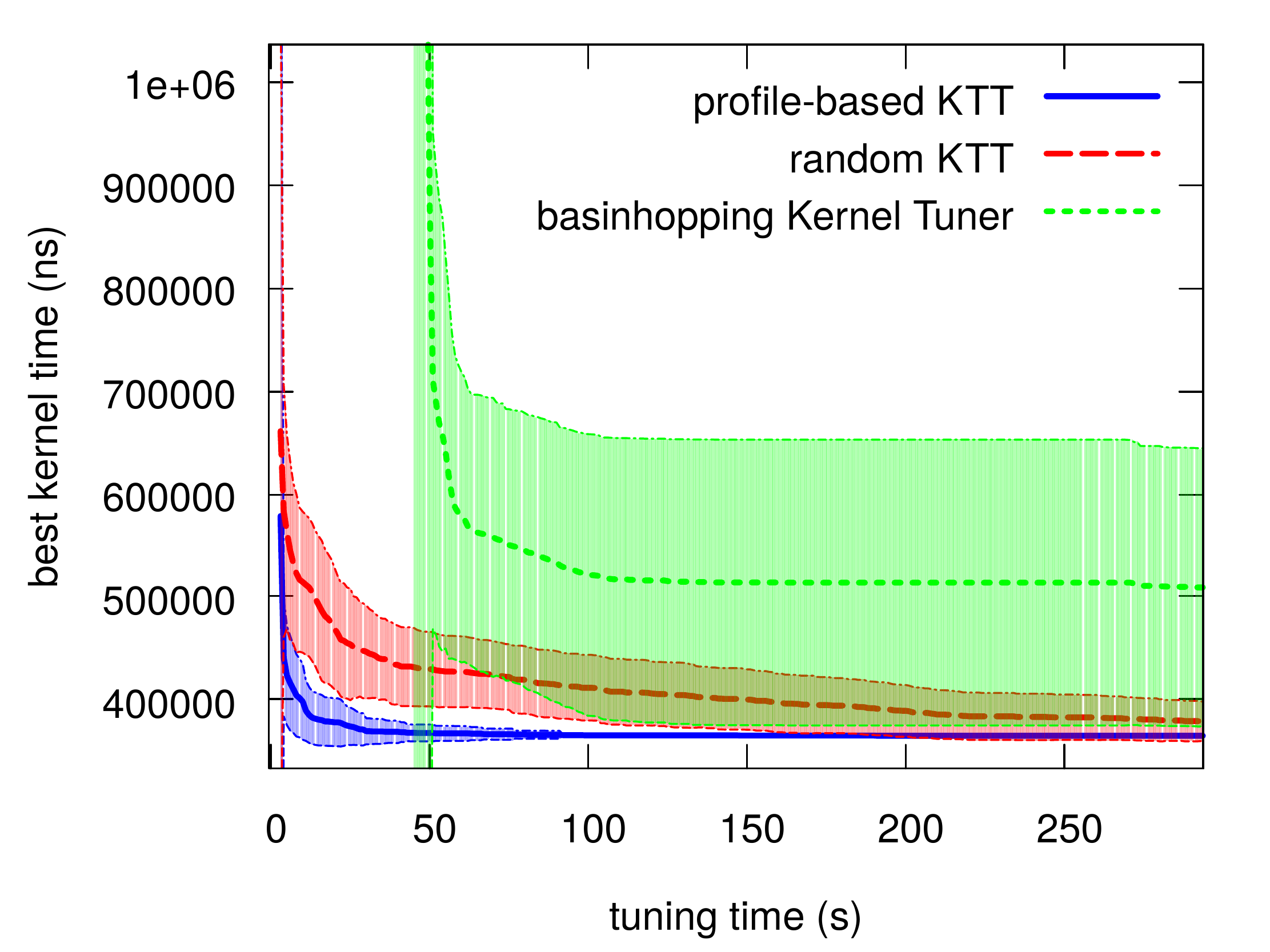}
  \includegraphics[width=.49\hsize]{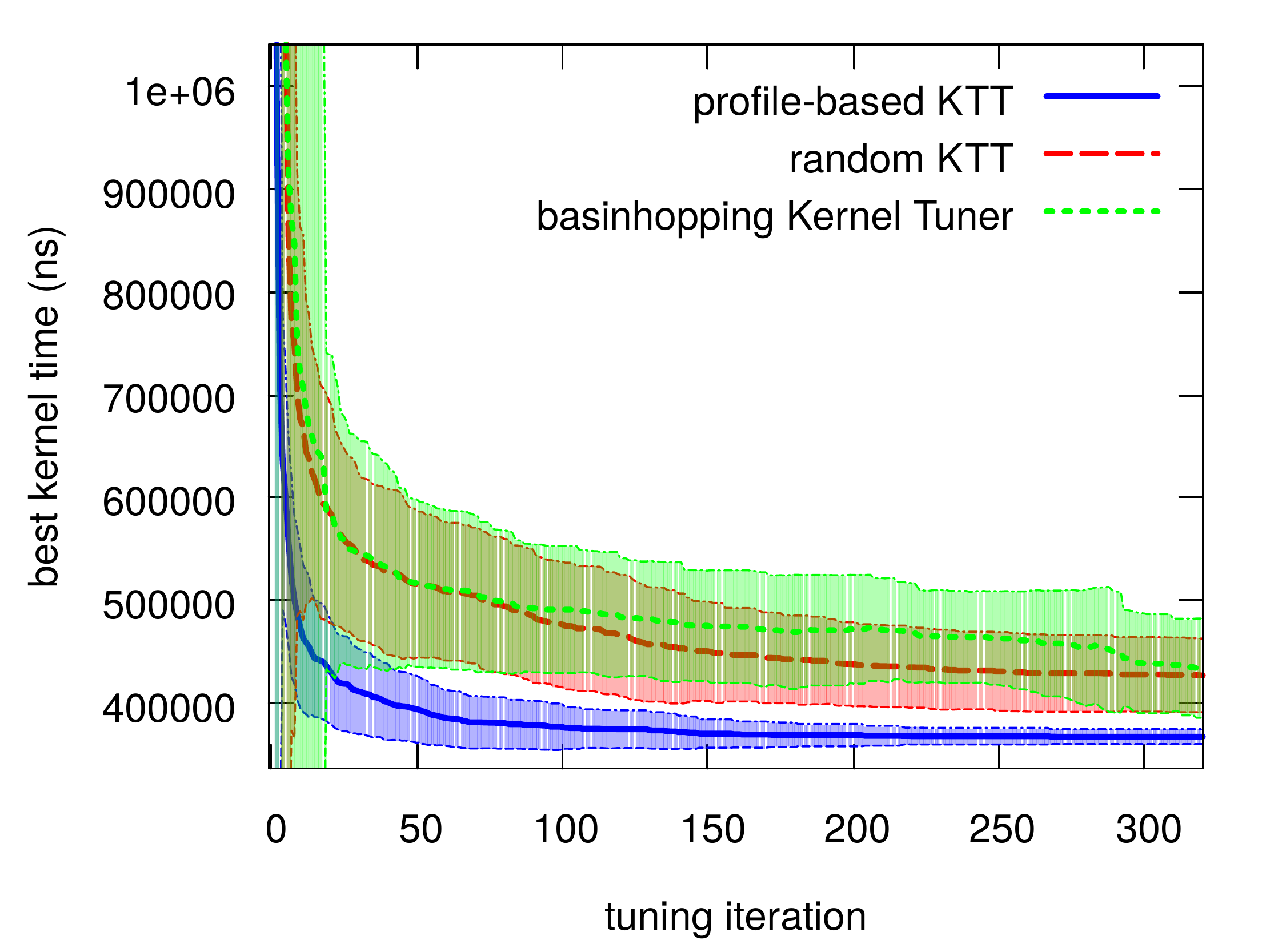}
  \caption{Convergence of the Convolution benchmark using KTT and Kernel Tuner. Left: convergence speed in time. Right: comparison of iterations (empirical tests). The solid line shows the average, transparent area shows the standard deviation.}
  \label{fig:conv_kt}
\end{figure}

Similarly to n-body, the Basin Hopping method converges slowly in the Convolution benchmark, see Figure~\ref{fig:nbody_kt}. Comparing the number of empirical tests, it works similarly to the random searcher. However, it converges much slower when real-time tuning is measured. 

Both the Matrix transposition and Convolution benchmarks reduce tuning space significantly using pre-defined constraints. When tuning space (created as the cross product of pre-defined tuning parameter values) is pruned by constraints, only 1.55\% of configurations remain with Matrix transposition, and 0.025\% of configurations remain with Convolution. We speculate that this may cause a significant delay at the start of tuning (16 seconds with Matrix transposition and 45 seconds with Convolution). In contrast, 41.7\% of configurations remain after constraint application with Coulomb sum, 33.3\% with n-body and 22.1\% with GEMM. Although this issue could be solved by optimizing the Kernel Tuner code, the comparison of empirical tests still shows Basin Hopping needs more empirical tests than the proposed searcher.

\subsection{Comparison to regression trees}
\label{sect:starchart}

To compare our approach to regression trees, we employed Starchart~\cite{jia2013starchart}, which trains regression trees to predict the outcome variable (usually performance or consumed power) based on the values of tuning parameters. With the trained tree, one can predict the outcome for the entire space and find the configurations with the best predicted outcomes. 

\subsubsection{Experiment setup} 

In their evaluation, Starchart authors started the training with 20 datapoints and added more iteratively until prediction accuracy, measured as the median relative prediction error, got below 15\%, or a maximum of 200 training datapoints was reached. Prediction accuracy was evaluated on 200 validation datapoints. Both training and validation datapoints were selected by uniform random sampling. It is worth noting that the tuning spaces used to evaluate Starchart in~\cite{jia2013starchart} were often very large as they contained entire ranges of values instead of only a few meaningful ones, \eg{}, 32-1024 instead of just 32, 64, 128, 256, 512 and 1024 as block sizes.



We evaluate the accuracy of the approach as follows:
\begin{itemize} 
  \item we randomly select 200 tuning configurations as a testing set;
  \item we perform training by measuring additional tuning configurations until median relative prediction error gets below 15\%;
  \item we measure how many configurations, sorted by best predicted kernel time, need to be examined to find one that is actually good. 
\end{itemize}
In the last step of the protocol, we take the whole tuning space, calculate the predictions and sort them from the best to the worst. In this order, we go one by one and examine whether the configuration is actually well-performing, \ie{}, within 110\% of the best kernel time. 

We have evaluated five benchmarks (coulomb, gemm-reduced, conv, mtran, nbody) using GeForce RTX 2080. 
 Because Starchart is not a tuning toolkit, we cannot directly compare tuning speed in time. Therefore, we compare the number of empirical tests here.

\subsubsection{Results}

\begin{table}
\caption{Results of autotuning using Starchart compared to the random searcher, using GeForce GTX 1070 and GeForce RTX 2080. All columns show the number of empirical tuning steps. The \textit{model build} column shows steps required for model training and testing, \textit{tuning} shows steps required for tuning before a well-performing configuration is found, and \textit{random} shows the number of steps required by random searcher (taken from Table~\ref{tab:random_iters}).}
\small
\centering
\subcaption*{GeForce GTX 1070}
\hspace{0.4cm}\begin{tabular}{|l|l|l|l|l|}
\hline
	        & model build 	& tuning 	& random \\ \hline
Coulomb sum  	& 233  		& 26	 	& 33\\
Matrix trans.  	& 378 		& 14	 	& 9\\
GEMM	  	& 400 		& 636	 	& 449\\
n-body  	& 397 		& 21	 	& 36\\
Convolution  	& 321 		& 278	 	& 348\\
\hline
\end{tabular}
\newline
\centering
\subcaption*{GeForce RTX 2080}
\begin{tabular}{|l|l|l|l|l|}
\hline
	        & model build 	& tuning 	& random \\ \hline
Coulomb sum  	& 234  		& 7	 	& 15\\
Matrix trans.  	& 397 		& 50	 	& 46\\
GEMM	  	& 391 		& 413	 	& 259\\
n-body  	& 369 		& 9	 	& 38\\
Convolution  	& 259 		& 9	 	& 567\\
\hline
\end{tabular}
\label{tab:pc_starchart10702080}
\end{table}

We performed the comparison of Starchart and random searcher for GeForce GTX 1070 and for GeForce RTX 2080 in Table~\ref{tab:pc_starchart10702080}. The tables show the number of tuning steps required to build a model and to perform a search until a well-performing configuration is found. The values for random searcher are taken from Table~\ref{tab:random_iters}. As can be seen, including the model build phase, Starchart performs worse than the random searcher, excepting the Convolution benchmark on GeForce RTX 2080. We believe that a well-designed tuning space, which does not contain a vast amount of poorly-performing tuning configurations, cannot be easily tuned with surrogate models created as part of the tuning process itself.

\begin{table}[ht]
\centering
\small
\begin{tabular}{|l|l|l|}
\hline
	        & SC@1070 	& proposed@1070	\\ \hline
Coulomb sum  	& 3	 	& 5		\\
Matrix trans.  	& 42	 	& 18		\\
GEMM	  	& 564	 	& 16		\\				
n-body  	& 17	 	& 6		\\
Convolution  	& 9	 	& 26		\\
\hline
\end{tabular}
\caption{Results of autotuning using Starchart compared to the proposed searcher, using GeForce RTX 2080 and models trained on GeForce GTX 1070. Both columns show the number of empirical tuning steps.}
\label{tab:pc_starchart_portability}
\end{table}

Although Starchart is not designed to ensure model portability (\ie{}, using the model trained on a different GPU), we have performed a test when a regression tree from a different GPU is used. The idea behind this experiment is that if performance portability is good for some of the well-performing configurations (\ie{}, there are configurations which perform well on both GPUs), the regression tree trained on a different GPU can be used successfully. We build the regression tree on GeForce GTX 1070 and use it to navigate searching on GeForce RTX 2080. We have compared the Starchart results to our searcher using a model from GeForce GTX 1070 to navigate tuning on GeForce RTX 2080. The comparison is given in Table~\ref{tab:pc_starchart_portability}. In this scenario, Starchart can compete with the proposed searcher in case the regression tree can describe the tuning space well, and some of the well-performing configurations of the GPU used for model build are also well-performing on the GPU used for autotuning. However, our model does not have such a restriction. The proposed searcher also seems to be more robust: there are no long tuning times as with the Starchart and GEMM example (on both GeForce GTX 1070 and RTX 2080) and Convolution (on GeForce GTX 1070).


\section{Related Work}
\label{sect:related_work}
Tuning space search methods can be divided into three categories: (i) model-free methods, which are viewing tuning space search as a mathematical optimization problem and empirically search for the best configuration; (ii) model-based methods, which use a performance or power model to select the best configuration from the tuning space directly; (iii) hybrid methods, which combine empirical search with a model.

In this section, we compare our approach with the state-of-the-art methods in tuning space search, and also with methods using historical profiling data and performance counters for code optimization.

\subsection{Model-free methods}
The model-free methods based on optimization are used in the majority of code optimization parameter autotuning frameworks~\cite{ansel2014opentuner, gerndt2015automatic, nugteren2015cltune, rasch2018atf, vanwerkhoven2018kernel}. As tuning spaces are not continuous, contain boolean tuning variables and are non-linear and non-convex,  searching tuning spaces is challenging. Therefore, in some experiments, the random search seems more reliable than more sophisticated search methods~\cite{seymour2008comparison, nugteren2015cltune}. On the other hand, some authors show improvement over random search. In~\cite{balaprakash2011can}, the Nelder-Mead method has been successfully adapted to non-continuous tuning spaces. More recently, it has been shown that the combination of global and local search has the potential to outperform random search~\cite{vanwerkhoven2018kernel} systematically. The difference between model-free optimization-based searching and our method is that optimization has to be executed from scratch when hardware or a performance-relevant input parameter changes, whereas our method uses a model portable across different GPUs and inputs. In this paper, we compare the results of our searcher to~\cite{vanwerkhoven2018kernel}. We confirmed result from~\cite{vanwerkhoven2018kernel} showing that Basin Hopping coupled with local search is superior to random search. We also show that our searcher outperforms Basin Hopping in both number of empirical tests and convergence time.

\subsection{Model-based methods}
The model-based methods need to use a model which can be evaluated faster than empirical testing. Therefore, simulators such as GPGPU-Sim~\cite{bakhoda2009analyzing} are not practical for this purpose. Multiple models for analysis and performance prediction of CUDA or OpenCL code have been developed~\cite{hong2009analytical, baghsorkhi2010adaptive, zhang2011quantitative, sim2012performance, huang2014gpumech, li2015transit}. However, the construction of those models often needs manual effort, which makes them impractical for autotuning. Moreover, independent studies reveal high inaccuracies in their ability to predict performance on some problems~\cite{volkov2016understanding, madougou2016landscape}, so it is questionable whether they can be used to select good tuning configurations from a vast tuning space. On the other hand, the model-based tools seem to have promising results if only one tuning parameter is searched. For example, it has been shown that a model can replace empirical tuning to select a well-performing work-group size (thread block in CUDA terminology)~\cite{cummins2017end, connors2019automatically, yu2020efficient}, thread-coarsening factor~\cite{cummins2017end} or code variant~\cite{muralidharan2016architecture}. In our work, we focus on the optimization of many tuning parameters together.

\subsection{Hybrid methods}
Hybrid methods introduce a surrogate model which is not intended to find the best tuning configuration directly, but rather to prune or bias empirical searching. In~\cite{falch2015machine}, a machine learning model is built from a sample of tuning space, and only the tuning configurations with the best predicted performance are empirically tested. The authors show that their method can outperform the random search. Regression trees have been used to speed up autotuning in multiple studies~\cite{feng2017sampling, jia2013starchart, price2015improving}. The regression trees are built from a representative sample of the tuning space, and their precision can be improved during search~\cite{feng2017sampling}. All the papers evaluate their approach using rather vast tuning spaces, \eg{}, testing all integer thread block sizes in the interval $<1, 1024>$, instead of using more rational sizes $2^n, n \in <5, 10>$. Therefore, the regression trees can be built from a tiny fraction of the tuning space. With rationally constructed tuning spaces, such as those presented in~\cite{nugteren2015cltune, petrovic2020benchmark}, a large portion of tuning space has to be used for tree construction, as we show experimentally in Section~\ref{sect:starchart}. As the models have to be re-trained when hardware or performance-relevant input characteristics change, they are not practical when a large portion of tuning space is needed for training. In contrast, our approach allows for the use of a model trained on different hardware or input.

\subsection{Methods using historical data}
Only a few works use historical autotuning data (\eg{}, obtained on different hardware) to improve autotuning convergence speed. In~\cite{muralidharan2016architecture}, authors predict the best code variant of a kernel for unseen GPU by using data obtained on a set of explored GPUs. The difference in our work is that we focus on searching a multi-dimensional tuning space, and only one GPU is sufficient for empirical exploration with our model. The historical data observed on one CPU architecture are used to bias a multi-dimensional search on a different CPU architecture in~\cite{roy2016exploiting}. Their approach works well if the fast implementations on both architectures are correlated. Our approach does not require such a correlation. A model which automatically selects a tuning configuration for a specific application input is presented in~\cite{tillet2017input}. It is trained on different inputs and then generalizes how to select the right tuning configuration for the used input. Compared to our model, it does not need to repeat empirical search when the input changes. On the other hand, our model does not need to be trained on multiple inputs and GPUs. In our previous work~\cite{olha2020exploiting}, we used data measured on one hardware to prune dimensions on different hardware. While this approach works well for speeding up exhaustive search, it brings no advantage when coupled with a searcher based on mathematical optimization.

\subsection{Methods using performance counters}
The typical usage of hardware performance counters is to profile a kernel for manual inspection of bottlenecks. However, there are several works which use performance counters to navigate optimization automatically. 

In~\cite{konstantinidis2015practical, wu2015gpgpu}, performance counters are used to approximate the performance and scaling of a known implementation on unseen GPUs. In contrast, our work focuses on navigating tuning space search, \ie{}, changing the implementation to maximize its performance. 

In~\cite{muralidharan2016architecture}, authors use performance counters to detect relevant features for their code variant selection model. Their model is trained on multiple code variants (alternative functionally equivalent implementations) and multiple GPUs. It predicts the fastest code variant on unseen GPU. They optimize one-dimensional tuning space and require multiple GPU architectures for training, whereas our approach allows searching many-dimensional tuning spaces after training on one GPU.

CPU performance counters have been used to select a well-performing combination of the number of OpenMP threads and scheduling strategy~\cite{wang2009mapping}, power-efficient process scheduling~\cite{singh2009prediction}, a well-performing combination of compiler optimization switches~\cite{cavazos2007rapidly}, a well-performing combination of activated or deactivated prefetchers~\cite{rahman2015maximizing}, and well-performing settings of stream configuration on Xeon Phi~\cite{zhang2018auto}. The main difference between the approach presented in this paper is that we can tune any user-defined optimization (transformation of the source code), whereas~\cite{wang2009mapping, singh2009prediction, cavazos2007rapidly, rahman2015maximizing, zhang2018auto} are developed to perform a specific type of optimization (\eg{}, setting OpenMP scheduling strategy and number of threads). With fixed optimizations, end-to-end learning is possible: the input of the machine-learning algorithm is a vector of tuning parameters' values, performance counters, and runtimes obtained from multiple applications, each running at different processors with various inputs. The output can be predicted runtime, or the predicted values of the best tuning parameters. Our method allows each tuned application to use original tuning parameters, so such a training set cannot be constructed. Therefore, we combine the machine learning model (created for each tuned application separately) with the expert system, navigating tuning space search according to predicted $PC_{ops}$ instead of predicted runtimes or best tuning parameters' values. 


\section{Conclusion and future work}
\label{sect:conclusion}
In this paper, we have introduced a novel tuning space searcher, which uses hardware performance counters to speed up the searching process. The proposed profile-based searcher builds a model of relations between tuning parameters and performance counters using any GPU and input, and uses this model to navigate searching on an unseen GPU or input. We have experimentally shown that the performance counters can significantly reduce the number of empirical tests that have to be performed in autotuning. We have also shown that the real-time convergence of the proposed searcher is typically superior to random searcher or state-of-the-art optimization-based Basin Hopping searcher implemented in Kernel Tuner~\cite{vanwerkhoven2018kernel}. The proposed searcher is implemented in Kernel Tuning Toolkit~\cite{petrovic2020benchmark}, and therefore it can be used for both offline and dynamic autotuning.

Although the searcher was designed for CUDA-enabled GPUs, we believe that a similar searcher can be developed for GPUs of different vendors or even different processor architectures, such as CPUs.

In future work, we plan to improve the searcher performance. As discussed in Section~\ref{sect:limits}, some components of the proposed profile-based searcher can be replaced by a more sophisticated implementation. We plan to experiment with (potentially gradient-based) local search methods, predict performance counters for the hardware used for autotuning and replace the system for bottleneck analysis and reaction with a machine-learning model.

Apart from improving the proposed searcher, we plan to investigate other possibilities to leverage hardware performance counters. First, we plan to extend our system to predict how well-tuned the actual configuration of a kernel is. Such a prediction allows us to stop the tuning at the right time, as well as predict how much performance can be gained by autotuning (which is especially important during dynamic autotuning, when the tuning time is limited). Second, we plan to use the vast amount of tuning data for studying the behavior and efficiency of different HW architectures and code optimization strategies.

The source code of Kernel Tuning Toolkit, the proposed profile-based searcher, and profiled data conducted for this study are available to the community. Therefore, it is easy to replicate results from this study, test the searcher with different benchmarks, or modify or extend our searcher implementation.

\section*{Acknowledgements}
\small
The work was supported from European Regional Development Fund-Project "CERIT Scientific Cloud" (No. CZ.02.1.01\-/0.0/0.0/16\_013/0001802). Computational resources were supplied by the project "e-Infrastruktura CZ" (e-INFRA LM2018140) provided within the program Projects of Large Research, Development and Innovations Infrastructures.

\bibliographystyle{plain}
\bibliography{fila}

\end{document}